\PassOptionsToPackage{hyphens}{url}

\documentclass[prl,aps,reprint,amsmath,amssymb,longbibliography, superscriptaddress]{revtex4-2}

\PassOptionsToPackage{bbgreekl}{mathbbol}

\DeclareUnicodeCharacter{0304}{={}} 

\usepackage[]{ajtex-revtex}

\renewcommand*{\arraystretch}{1.2}
\usepackage{multirow}

\makeatletter 
    
\renewcommand\onecolumngrid{
\do@columngrid{one}{\@ne}%
\def\set@footnotewidth{\onecolumngrid}
\def\footnoterule{\kern-6pt\hrule width 1.5in\kern6pt}%
}

\renewcommand\twocolumngrid{
\def\footnoterule{
\dimen@\skip\footins\divide\dimen@\thr@@
\kern-\dimen@\hrule width.5in\kern\dimen@}
\do@columngrid{mlt}{\tw@}
}%

\makeatother

\makeatletter
\def\l@subsubsection#1#2{}
\makeatother

\renewcommand{\paragraph}[1]{\vspace{.5em}\noindent\textbf{#1}}

\begin{document} 

\title{Thermodynamics of ideal spin fluids and pseudo-gauge ambiguity}

\author{Jay Armas}\email{j.armas@uva.nl}
\affiliation{Institute for Theoretical Physics, University of Amsterdam, 1090 GL Amsterdam, The Netherlands}
\affiliation{Dutch Institute for Emergent Phenomena, 1090 GL Amsterdam, The Netherlands}
\affiliation{Institute for Advanced Study, University of Amsterdam, Oude Turfmarkt 147, 1012 GC Amsterdam, The Netherlands}
\affiliation{Niels Bohr International Academy, The Niels Bohr Institute, University of Copenhagen,
Blegdamsvej 17, DK-2100 Copenhagen \O{}, Denmark}

\author{Akash Jain}\email{akash.jain@maths.ox.ac.uk}
\affiliation{Mathematical Institute, University of Oxford, Woodstock Road, Oxford, OX2 6GG, U.K.}

\date{\today}

\begin{abstract}
Conserved currents of relativistic spin fluids derived from microscopic models appear to violate local thermodynamic relations. We present a systematic analysis of pseudo-gauge improvements in ideal spin hydrodynamics and identify a family of pseudo-gauges where standard thermodynamic relations are satisfied. We quantify pseudo-gauge ambiguities in the spin equation of state and derive universal thermodynamic relations that apply to conserved currents in any pseudo-gauge. As an application, we extract the thermodynamic variables and equations of state for free Dirac fermions and scalar fields.
\end{abstract} 

\pacs{Valid PACS appear here}

\maketitle

Relativistic heavy-ion collisions produce a short-lived quark-gluon plasma (QGP) that is well described by relativistic hydrodynamics \cite{Teaney:2003kp, Romatschke:2007mq}. The recent observation of global $\Lambda$-hyperon polarization by the STAR Collaboration demonstrated that the QGP fireball carries substantial angular momentum and exhibits strong vortical structure \cite{STAR:2017ckg, 2019arXiv190511917S-short, 2020PhRvC.101d4611A-short, 2022PhRvL.128q2005A-short}.
This discovery has motivated the development of \emph{spin hydrodynamics}, a theoretical framework that describes collective spin phenomena and provides novel probes of the QGP’s gradients and rotational dynamics~\cite{2011PPNL....8..801B, Florkowski:2017ruc, Florkowski:2018fap, Montenegro:2017rbu, Hattori:2019lfp, Florkowski:2019qdp, Becattini:2020ngo, Singh:2020rht, Fukushima:2020ucl, Hongo:2021ona, Gallegos:2021bzp, Florkowski:2021wvk, Gallegos:2022jow, Peng:2021ago, Cao:2022aku, Weickgenannt:2022zxs, 2022PhRvD.106i1901W, Biswas:2023qsw, Florkowski:2024cif, Huang:2024ffg, Singh:2025hnb, Daher:2025pfq}.

Hydrodynamics is a theory of conserved currents. However, these currents admit improvements—the so-called pseudo-gauge transformations—that leave the global conserved charges unchanged~\cite{halbwachs1960théorie, Hehl:1976vr, 2011PhRvD..84b5013B, Speranza:2020ilk, Becattini:2025twu}. As a result, the conserved currents in spin hydrodynamics are not unique, even in equilibrium. In particular, the constitutive relations in generic pseudo-gauges fail to satisfy the standard thermodynamic relations implied by statistical mechanics \cite{Becattini:2023ouz, Becattini:2025oyi, 2025PhRvL.134h2302F, Drogosz:2024gzv}. 
See also~\cite{2025PhRvL.134h2302F, Drogosz:2024gzv} for generalised vector thermodynamic relations for ideal spin fluids, which we will not delve into in this work.
This subtlety prevents a direct identification of thermodynamic variables of spin hydrodynamics through naive comparison with the conserved currents obtained from microscopic models or experimental measurements.

For instance, the particle current admits a pseudo-gauge ambiguity $J^\mu \to J^\mu + \dow_\sigma M^{\sigma\mu}_J$, for some antisymmetric tensor $M^{\mu\nu}_J$. This shifts the particle density in the fluid's rest frame as $n \to n - u_\mu \dow_\sigma M^{\sigma\mu}_J$, where $u^\mu$ is the fluid four-velocity. Which of these is the correct thermodynamic density? This issue does not arise in ordinary hydrodynamics because all fluid variables are constant in equilibrium, leading to the vanishing of such improvements. Conversely, fluid variables admit a non-trivial spacetime profile in spin hydrodynamics. Concretely, taking $M_J^{\mu\nu} = f(T)\dow^{[\mu}(u^{\nu]}/T)$, where $f(T)$ is an arbitrary function of the fluid temperature $T$, the particle density shifts to $n \to n - f'(T) a^2$, where $a^2$ is the squared fluid acceleration. This ambiguity means that $n$ may violate standard thermodynamic relations in a generic pseudo-gauge.

In this work, we perform an exhaustive study of the effect of pseudo-gauge improvements on ideal spin hydrodynamics. We identify a family of \emph{thermodynamic pseudo-gauges}, where the conserved currents obey standard thermodynamic relations. Working perturbatively in spin, we derive robust \emph{thermodynamic relations} that the conserved currents in any pseudo-gauge must satisfy to be compatible with thermodynamics. We use these results to derive the spin-sensitive equation of state (EoS) and thermodynamic variables for free Dirac fermions and scalar fields. This allows us to resolve the apparent mismatch in the literature between spin hydrodynamics with local thermodynamic relations~\cite{Becattini:2023ouz, Becattini:2025oyi, 2025PhRvL.134h2302F, Drogosz:2024gzv}.

We demonstrate that the thermodynamic variables obtained through this procedure generically inherit pseudo-gauge ambiguities themselves. A notable exception arises in conformal (scale-invariant) theories—such as massless free fields—where the stronger symmetry constraints eliminate this ambiguity, allowing the thermodynamic variables to be identified unambiguously.


\paragraph{Spin hydrodynamics}.---Spin hydrodynamics consists of a non-symmetric energy-momentum tensor $T^{\mu\nu}$, spin current $\Sigma^{\mu\nu\rho}=\Sigma^{\mu[\nu\rho]}$, and possibly particle/Baryon current $J^\mu$. The respective conservation equations are
\begin{align}\label{eq:conservation}
    \dow_\mu T^{\mu\nu} = 0, \quad 
    \dow_\mu \Sigma^{\mu\nu\rho} = 2T^{[\nu\rho]}, \quad
    \dow_\mu J^\mu = 0.
\end{align}
These govern the dynamics of fluid variables: four-velocity $u^\mu$ (s.t. $u^\mu u_\mu = -1$), temperature $T$, spin chemical potential $\mu_{\mu\nu}=\mu_{[\mu\nu]}$, and particle chemical potential $\mu$. 

Focusing on 3+1 spacetime dimensions, it is helpful to identify a basis of spacelike vectors
\begin{equation}
    a^\mu \equiv \mu^{\mu\nu} u_\nu, \quad 
    \omega^\mu \equiv \half \epsilon^{\nu\mu\rho\sigma} u_\nu \mu_{\rho\sigma}, \quad
    \ell^\mu
    = \epsilon^{\nu\mu\rho\sigma} u_\nu a_\rho \omega_\sigma,
    \label{eq:basis-vectors}
\end{equation}
where we have used $\eta_{\mu\nu}=\diag(-1,1,1,1)$ and the Levi-Civita symbol is defined via $\epsilon_{0123}=1$. We also define the thermal-scaled quantities
\begin{gather}
    \beta^\mu \equiv \frac{u^\mu}{T}, \quad 
    \nu_{\mu\nu}\equiv\frac{\mu_{\mu\nu}}{T}, \quad 
    \nu \equiv \frac{\mu}{T}, \nn\\
    \alpha^\mu \equiv \frac{a^\mu}{T}, \quad 
    w^\mu \equiv \frac{\omega^\mu}{T}, \quad 
    l^\mu \equiv \frac{\ell^\mu}{T^2}.
    \label{eq:scaled-variables}
\end{gather}

Hydrodynamics is characterised by the constitutive relations for the currents in terms of the hydrodynamic variables. The constitutive relations must satisfy the local second law of thermodynamics~\cite{landau1959fluid}: there must exist an entropy current $S^\mu$ such that $\dow_\mu S^\mu \geq 0$ for all fluid configurations. We define the free energy current
\begin{align}
    N^\mu = S^\mu 
    + T^{\mu\nu} \beta_\nu 
    + \half \Sigma^{\mu\nu\rho}\nu_{\nu\rho}
    + J^\mu \nu,
\end{align}
in terms of which, the local second law becomes
\begin{equation}\label{eq:adiabaticity}
    \dow_\mu N^\mu
    \geq T^{\mu\nu} 
    \lb \dow_\mu\beta_\nu + \nu_{\mu\nu} \rb
    - \half\Sigma^{\mu\nu\rho}\dow_\mu\nu_{\nu\rho}
    - J^\mu \dow_\mu\nu.
\end{equation}
This requirement provides non-trivial constraints on the structural form of the constitutive relations; see~\cite{Hattori:2019lfp, Gallegos:2021bzp}.

A corollary of the local second law is that hydrodynamic equations admit an equilibrium solution~\cite{Bhattacharyya:2013lha}
\begin{equation}
    \beta^\mu\big|_\eqb 
    = \beta^\mu_0 + \nu_0^{\mu\nu}x_\nu,~~
    \nu_{\mu\nu}\big|_\eqb 
    = \nu_{0\mu\nu},~~
    \nu\big|_\eqb  = \nu_0,
    \label{eq:eqb-condition}
\end{equation}
where $\beta^\mu_0$, $\nu_{0\mu\nu}$, and $\nu_0$ are constant parameters characterising the thermal state~\footnote{One may redefine hydrodynamic variables, known as hydrodynamic frame transformations. \Cref{eq:eqb-condition} holds in the so-called thermodynamic frames. More generally, such as in Landau or Eckart frames, the equilibrium solution would be an appropriate redefinition of \cref{eq:eqb-condition}.}. 
In equilibrium, $a_\mu$ aligns with the fluid acceleration $u^\nu\dow_\mu u^\mu$, while $\omega^\mu$ aligns with the fluid vorticity $\half \epsilon^{\mu\nu\rho\sigma} u_\nu \dow_\rho u_\sigma$.

\paragraph{Pseudo-gauge ambiguity}.---%
Conserved currents, and hence the constitutive relations of hydrodynamics, are only defined up to improvements terms that leave \cref{eq:conservation} invariant. In spin hydrodynamics, this is known as the pseudo-gauge ambiguity and takes the form~\cite{halbwachs1960théorie, Hehl:1976vr, 2011PhRvD..84b5013B, Speranza:2020ilk}
\begin{align}\label{eq:pseudo-gauge}
    T^{\mu\nu}
    &\to T^{\mu\nu} 
    + \half\dow_{\sigma} \Big( 
    M_T^{\mu\nu\sigma}
    + M_T^{\nu\mu\sigma}
    + M_T^{\sigma\mu\nu}
    \Big), \nn\\
    \Sigma^{\mu\nu\rho}
    &\to \Sigma^{\mu\nu\rho}
    + M_T^{\mu\nu\rho}
    + \dow_\sigma M_\Sigma^{\sigma\mu\nu\rho}, \nn\\
    J^\mu 
    &\to J^\mu + \dow_\sigma M_J^{\sigma\mu},
\end{align}
for arbitrary tensors \smash{$M_J^{\mu\nu} = M_J^{[\mu\nu]}$}, \smash{$M_T^{\mu\nu\rho} = M_T^{\mu[\nu\rho]}$}, and
\smash{$M_\Sigma^{\mu\nu\rho\sigma} = M_\Sigma^{[\mu\nu][\rho\sigma]}$}.
There is a similar pseudo-gauge ambiguity $M_S^{\mu\nu}$ in the entropy current~\cite{Bhattacharyya:2012nq, Becattini:2023ouz}.

Pseudo-gauge ambiguity implies that the spin current has no independent information of its own and can always be set to zero with a suitable choice of $M_T$. Therefore, it is helpful to define the Belinfante tensor~\cite{belinfante1940current}
\begin{align}\label{eq:belinfante-tensor}
    T^{\mu\nu}_\rmB
    &\equiv T^{\mu\nu} 
    - \half\dow_{\rho} \lb \Sigma^{\mu\nu\rho}
    + \Sigma^{\nu\mu\rho}
    + \Sigma^{\rho\mu\nu}
    \rb,
\end{align}
which is invariant under $M_T$. However, it still transforms under $M_\Sigma$ as
\begin{align}
    T^{\mu\nu}_\rmB
    &\to T^{\mu\nu}_\rmB
    - \dow_\sigma\dow_{\rho} M_\Sigma^{\sigma(\mu\nu)\rho}.
\end{align}

\paragraph{Ideal spin fluids}.---Ideal spin fluids are a solution of the adiabaticity equation with the free energy current~\cite{Gallegos:2021bzp}
\begin{align}\label{eq:ideal-free-energy}
    N^\mu = p\,\beta^\mu~.
\end{align}
Here $p$ is the thermodynamic pressure of the fluid. We use its variation to obtain other thermodynamic quantities
\begin{align}
    \df p
    &= s \df T
    + \frac{1}{2} \rho^{\mu\nu} \df \mu_{\mu\nu}
    + \pi_\mu \df u^{\mu}
    + n \df\mu 
    , \nn\\
    \epsilon
    &= Ts + \half \mu_{\mu\nu}\rho^{\mu\nu} 
    + \mu n
    - p.
    \label{eq:spin-thermo}
\end{align}
We may identify the entropy density $s$, spin density $\rho^{\mu\nu}$, particle density $n$, and energy density $\epsilon$. As we are expressing the variation $\df p$ in terms of a tensor $\df\mu_{\mu\nu}$, Lorentz symmetry requires an additional term $\pi_\mu \df u^{\mu}$, such that 
$\rho^{[\mu}{}_{\!\rho}\mu^{\nu]\rho}=u^{[\mu}\pi^{\nu]}$ and $\pi_\mu u^\mu = 0$. See~\cref{app:thermo} for more details. We may interpret $\pi_\mu$ as the momentum density of the fluid in its rest frame.

We can express the EoS for $p$ as a function of the thermodynamic scalars: $T$, $\mu$, $a^2$, $\omega^2$, and $a\cdot\omega$. Identifying the acceleration, rotation, and mixed components of spin susceptibility, respectively
\begin{align}
    \chi_{aa} = 2\frac{\dow p}{\dow a^2}, \quad 
    \chi_{\omega\omega} = 2\frac{\dow p}{\dow\omega^2}, \quad     \chi_{a\omega} = \frac{\dow p}{\dow(a\cdot\omega)},
\end{align}
we can read off the spin and momentum densities
\begin{align}\label{eq:density-decomposition}
    \rho^{\mu\nu}
    &= 2\Big( \chi_{aa} a^{[\mu} 
    {+} \chi_{a\omega} \omega^{[\mu}\!\Big) u^{\nu]}
    + \epsilon^{\mu\nu\rho\sigma} u_\rho
    \Big( \chi_{a\omega} a_\sigma {+} \chi_{\omega\omega} \omega_\sigma \Big), \nn\\
    \pi^{\mu}
    &= 
    - \Big(\chi_{\omega\omega} + \chi_{aa} \Big) \ell^{\mu}.
\end{align}

Substituting the free energy current \eqref{eq:ideal-free-energy} into \cref{eq:adiabaticity}, we find the constitutive relations of an ideal spin fluid
\begin{align}\label{eq:ideal-fluid}
    T^{\mu\nu}
    &= \epsilon\,u^\mu u^\nu + p\,\Delta^{\mu\nu} 
    + u^{\mu} \pi^{\nu},
    \nn\\
    \Sigma^{\mu\nu\rho}
    &= \rho^{\nu\rho} u^\mu, \quad
    J^\mu 
    = n\,u^\mu, \quad
    S^\mu
    = s\, u^\mu,
\end{align}
where $\Delta^{\mu\nu} = \eta^{\mu\nu}+u^\mu u^\nu$.
It is straightforward to check that these are identically conserved on the equilibrium solution \eqref{eq:eqb-condition}. In general, \cref{eq:ideal-fluid} may be supplemented with non-equilibrium terms that vanish in equilibrium, which are not of interest to us in this work.
As one would expect, the thermodynamic densities align with their natural definitions in terms of the currents in \cref{eq:ideal-fluid}, i.e.
\begin{gather}
    \epsilon = u_\mu u_\nu T^{\mu\nu}, \quad 
    p = \frac13 \Delta_{\mu\nu} T^{\mu\nu}, \quad 
    \pi_\nu = -u_\mu \Delta_{\nu\rho} T^{\mu\rho}, \nn\\
    \rho^{\nu\rho} = - u_\mu S^{\mu\nu\rho}, \quad
    n = - u_\mu J^\mu, \quad 
    s = - u_\mu S^\mu.
\end{gather}
However, this delicate property is generally spoiled by pseudo-gauge improvements. 

The Belinfante tensor \eqref{eq:belinfante-tensor} for ideal spin fluids is
\begin{align}\label{eq:ideal-belinfante}
    T^{\mu\nu}_{\rmB}
    &=
    \epsilon_\rmB u^\mu u^\nu 
    + p_\rmB \Delta^{\mu\nu}
    - \chi_{aa} a^\mu a^\nu 
    + \chi_{\omega\omega} \omega^\mu \omega^\nu \nn\\
    &\qquad 
    + 2q_\rmB u^{(\mu}\ell^{\nu)},
\end{align}
where we have ignored non-equilibrium terms.
The new coefficients appearing above are given as
\begin{align}
    \epsilon_\rmB
    &= 
    \epsilon - 2(\chi_{aa}+\chi_{\omega\omega}) \omega^2
    - (a\cdot\omega)\df_T (T\chi_{a\omega}) \nn\\
    &\qquad 
    - a^2\df_T(T\chi_{aa})
    - \frac{\ell^2}{T^2}\df_A\chi_{aa},
    \nn\\
    p_\rmB
    &= p - \chi_{\omega\omega} \omega^2
    - \chi_{a\omega} (a\cdot\omega), \nn\\
    q_\rmB &=
    \frac{T}{2} \df_T\chi_{\omega\omega}
    + \frac{\omega^2}{2T^2}\df_A \chi_{\omega\omega} 
    + \frac{a\cdot\omega}{2T^2} \df_A \chi_{a\omega}.
\end{align}
We have introduced the notation $\df_\#$ for thermodynamic derivatives in the basis $\{T,\alpha^2,w^2,\alpha\cdot w,\nu\}$; refer to \cref{eq:scaled-variables}. We further define $\df_A = 2\df_{\alpha^2}+2\df_{w^2}$. We emphasise that the Belinfante energy and pressure are generically different from their thermodynamic counterparts.

\paragraph{Conformal spin fluids.}---The EoS for conformal (scale-invariant) spin fluids takes a restrictive form, such that $p/T^4$ is only a function of $\mu/T$ and $\omega^2/T^2$; see~\cite{Gallegos:2021bzp}. Consequently, we have
\begin{gather}
    p = T^4 f\lb\mu/T,\omega^2/T^2\rb, \quad 
    \epsilon = 3p, \quad 
    \chi_{aa} = \chi_{a\omega} = 0, \nn\\
    \epsilon_\rmB 
    = 3\epsilon - 2\omega^2\chi_{\omega\omega},~~
    p_\rmB = p - \omega^2\chi_{\omega\omega},~~
    q_\rmB = \frac{\dow(\omega^2\chi_{\omega\omega})}{\dow\omega^2}.
    \label{eq:conformal}
\end{gather}
The Belinfante tensor for conformal spin fluids, as we can check, is traceless. However, note that $\epsilon_\rmB \neq 3p_\rmB$.


\paragraph{Pseudo-gauge ambiguity of ideal spin fluids}.---%
The conserved currents derived from several microscopic models with spin are well known to deviate from the ideal spin-fluid forms in \cref{eq:ideal-fluid,eq:ideal-belinfante} \cite{Palermo:2021hlf, Palermo:2023ews, Ambrus:2014uqa, Ambrus:2019cvr, Ambrus:2019ayb, Buzzegoli:2017cqy}. In particular, the Belinfante tensor generally contains a nonvanishing $a^{(\mu}\omega^{\nu)}$ component, the particle current exhibits a nonzero $\ell^\mu$ term, while other components fail to satisfy the expected thermodynamic relations \cite{Becattini:2023ouz, Becattini:2025oyi, 2025PhRvL.134h2302F, Drogosz:2024gzv}.
We argue that the root cause of these discrepancies is that the comparison between microscopic calculations and hydrodynamics is not being performed in the appropriate pseudo-gauge.

The general Belinfante tensor, charge current, and entropy current for an ideal spin fluid takes the form
\begin{align}\label{eq:general-currents}
    T^{\mu\nu}_\rmB
    &= \cE u^\mu u^\nu 
    + \cP \Delta^{\mu\nu}
    \nn\\
    &\qquad 
    + \sum_{X,Y} \cX_{XY} X^\mu Y^\mu
    + 
    \sum_{X}
    2\cQ_{X} u^{(\mu}X^{\nu)}, \nn\\
    J^\mu 
    &= \cN u^\mu 
    + \sum_{X}
    \cN_X X^\mu, \quad 
    S^\mu 
    = \cS u^\mu 
    + \sum_{X}
    \cS_X X^\mu, 
\end{align}
where $X^\mu,Y^\mu\in \{a^\mu,\omega^\mu,\ell^\mu\}$ and $\cX_{XY}=\cX_{Y\!X}$. Let us also define $\cF = \cP + \omega^2 \cX_{\omega\omega}$ for later use. Since a symmetric 2-tensor can only have 10 independent components, we can set $\cX_{\ell\ell}=0$. Requiring that the currents are identically conserved in equilibrium gives several identities among the components, outlined in \cref{app:identities}. The explicit components for free Dirac and scalar fields are given in \cref{app:micro-examples-fermion,app:micro-examples-scalar}. 

General pseudo-gauge improvements of the Belinfante tensor contain 20 parameters, $\gamma_i$, while that of the charge and entropy currents contain 5 parameters each, $\lambda_i$ and $\lambda^S_i$. E.g., the subset of these that affect the currents till quartic order in spin for parity-preserving theories are
\begin{align}\label{eq:pseudo-gauge-params}
    \hspace{-.5em}
    M_\Sigma^{\mu\nu\rho\sigma}
    &= 4\gamma_1\, u^{[\mu} \Delta^{\nu][\rho}u^{\sigma]}
    + 2\gamma_2\, \Delta^{\mu[\rho}\Delta^{\sigma]\nu} \nn\\
    &\quad 
    + 4\gamma_7\, u^{[\mu}a^{\nu]} a^{[\rho} u^{\sigma]}
    + 4\gamma_8\, u^{[\mu}\omega^{\nu]} \omega^{[\rho} u^{\sigma]} 
    \nn\\
    &\quad 
    + 2\gamma_{13}\, 
    \lb u^{[\mu}a^{\nu]}\bar\mu^{\rho\sigma}
    + \bar\mu^{\mu\nu}u^{[\rho}a^{\sigma]}\rb
    + \cO(\mu_{\mu\nu}^4)
    , \nn\\
    M^{\sigma\mu}_J
    &= 2\lambda_1 u^{[\sigma} a^{\mu]}
    + \lambda_2 \bar\mu^{\sigma\mu} 
    + 2\lambda_5 a^{[\sigma} \ell^{\mu]} 
    + \cO(\mu_{\mu\nu}^5),
\end{align}
and similarly for $M_S^{\sigma\mu}$,
where $\bar\mu^{\mu\nu} = \Delta^{\mu\rho}\Delta^{\nu\sigma}\mu_{\rho\sigma}$.

Pseudo-gauge improvements shift the components in \cref{eq:general-currents}, denoted as $\cP' = \cP + \delta\cP$, etc. See \cref{app:counting,app:PG-current,app:PG-stress} for the explicit expressions. We want these to obey the thermodynamic form in \cref{eq:ideal-fluid,eq:ideal-belinfante}. This leads to the following constraints
\begin{gather} 
    \scD \cX'_{aa}
    = - 2\frac{\dow\cF'}{\dow a^2},~~
    \scD \cX_{\omega\omega}'
    = 2\frac{\dow\cF'}{\dow\omega^2},~~
    \cX_{a\omega,a\ell,\omega\ell}',\cQ_{a,\omega}' = 0, \nn\\
    \scD\cN'
    = \frac{\dow\cF'}{\dow\mu}, \quad
    \scD
    \cS'
    = \frac{\dow\cF'}{\dow T}, \quad
    \cN_{a,\omega,\ell}',~\cS_{a,\omega,\ell}'
    = 0,
    \label{eq:thermo-constraints}%
\end{gather}
where $\scD \equiv 1 - (a\cdot\omega)\dow/\dow(a\cdot\omega)$. There are also constraints involving $\cE'$, $\cQ_\ell'$ components, but they are identically satisfied due to the conservation identities; see \cref{app:details-thermo-constrains}.
\Cref{eq:thermo-constraints} provides linear partial differential equations (PDEs) for pseudo-gauge parameters $\gamma_i$'s and $\lambda_i$'s. Having solved these, the EoS of ideal spin fluids is given by the indefinite integral
\begin{align}\label{eq:EOS-formula}
    p
    &= -(a\cdot\omega)\int\frac{\df(a\cdot\omega)}{(a\cdot\omega)^2}\,\cF'.
\end{align}
The constant of integration vanishes for parity-preserving theories. We will specialise to the explicit examples of free Dirac fermions and scalar fields below.

\paragraph{Thermodynamic relations.}---%
The PDEs \eqref{eq:thermo-constraints} do not admit solutions for arbitrary conserved currents in \cref{eq:general-currents}. For solutions to exist, the currents must satisfy certain thermodynamic relations. In the absence of spin, we find the familiar relations: $\lim_{\mu_{\mu\nu}\to0}\cN - \dow\cP/\dow\mu=0$ and $\lim_{\mu_{\mu\nu}\to0}\cS - \dow\cP/\dow T=0$. Note that these do not follow from conservation. Performing an exhaustive analysis of pseudo-gauge improvements at quadratic order in spin (see \cref{app:details-thermo-constrains}), we find one thermodynamic relation for the Belinfante tensor and one for the particle current 
\begin{subequations}
\begin{align}
    \lim_{\mu_{\mu\nu}\to0}
    \lb \cX_{aa}
    + 2\frac{\dow\cF}{\dow a^2}
    \rb
    + \frac{T^2}{3}\df_T
    \lb \frac{\cX_{\omega\omega}}{T}
    - \frac2T \frac{\dow\cF}{\dow\omega^2} 
    \rb
    &= 0, \nn\\
    \lim_{\mu_{\mu\nu}\to0}\lb\frac{2\dow}{\dow a^2}
    - \df_T\frac{T\dow}{\dow\omega^2}\rb
    \lb\cN-\frac{\dow\cF}{\dow\mu}\rb
    - 2\cN_\ell 
    &= 0.
\end{align}
These relations are pseudo-gauge independent, i.e. they must be satisfied by the equilibrium conserved currents in any pseudo-gauge.
We find an analogous relation for the entropy current, with $\cN-\dow\cF/\dow\mu$ and $\cN_\ell$ replaced by $\cS-\dow\cF/\dow T$ and $\cS_\ell$ respectively. Similar thermodynamic relations exist at each order in spin. For instance, at the quartic order we find
\begin{widetext}
\begin{align}
    \lim_{\mu_{\mu\nu}\to 0}
    \lb \frac{2\dow}{\dow a^2}
    - \frac{1}{2T}\df_T\frac{T^2 \dow}{\dow\omega^2} 
    \rb \lb \cX_{aa}
    + 2\frac{\dow\cF}{\dow a^2} \rb
    + \lb \frac{T}{3}\df_T\frac{\dow}{\dow a^2}
    - \frac{1}{6T} \df_T^2\frac{T^3\dow}{\dow\omega^2}\rb
    \lb \cX_{\omega\omega} - 2\frac{\dow\cF}{\dow\omega^2} \rb
    - \frac{1}{3T^5} \df_T\frac{T^6\dow\cX_{a\omega}}{\dow(a\cdot\omega)}
    &= 0, \nn\\[0.8em]
    \lim_{\mu_{\mu\nu}\to 0}
    \lb \frac{4\dow^2}{\dow a^4}
    + \frac{1}{2T^4}\df_T
    \lb T^3\df_T\frac{T^3\dow^2}{\dow\omega^4}\rb
    - \frac{2}{T^2}\df_T
    \frac{T^3\dow^2}{\dow a^2\dow\omega^2} \rb
    \lb \cN - \frac{\dow\cF}{\dow\mu} \rb
    - \lb \frac{8\dow}{\dow a^2}
    - \frac{2}{T^2} \df_T
    \frac{T^3\dow}{\dow\omega^2}\rb \cN_\ell
    &= 0,
\end{align}
\end{widetext}
and similarly for the entropy current.
We have verified that these rather non-trivial thermodynamic relations hold for the special cases of equilibrium conserved currents of free Dirac fermions and scalar fields obtained in \cite{Palermo:2021hlf, Palermo:2023ews, Ambrus:2014uqa, Ambrus:2019cvr, Ambrus:2019ayb, Buzzegoli:2017cqy}. 
\label{eq:thermo-relations}
\end{subequations}

\paragraph{Pseudo-gauge-invariants.}---%
Having identified the conservation identities and thermodynamic relations, we wish to quantify genuine pseudo-gauge-independent information in equilibrium conserved currents. Clearly, the spin-free pressure, $p_0 \equiv \lim_{\mu_{\mu\nu}\to0} \cP$, is an invariant. We find exactly one such pseudo-gauge-invariant at quadratic and one at quartic order in spin
\begin{align}\label{eq:invariants-current}
    \cI_2
    &\equiv
    \lim_{\mu_{\mu\nu}\to 0}
    \frac{2\dow\cF}{\dow a^2}
    - T\df_T \frac{\dow\cF}{\dow\omega^2}, \nn\\
    \cI_4
    &\equiv
    \lim_{\mu_{\mu\nu}\to 0}\!
    \frac{4\dow^2\cF}{\dow a^4}
    + \lb \frac{1}{4T^3} \df_T 
    \frac{T^5\dow}{\dow\omega^2}
    - \frac{T\dow}{\dow a^2} \rb \df_T
    \frac{2\dow\cF}{\dow\omega^2}.
\end{align}
Evaluating these in a thermodynamic pseudo-gauge, we can relate the invariants to thermodynamic quantities such that
\begin{align}\label{eq:invariants}
    \cI_{2} 
    &= 
    \lim_{\mu_{\mu\nu}\to 0}
    \chi_{aa} 
    - \frac{T}{2}\df_T \chi_{\omega\omega}, \nn\\
    \cI_{4} 
    &=
    \lim_{\mu_{\mu\nu}\to 0}\!
    \frac{2\dow\chi_{aa}}{\dow a^2}
    + \lb \frac{1}{4T^3} \df_T 
    \frac{T^5\dow}{\dow\omega^2}
    - \frac{T\dow}{\dow a^2} \rb \df_T\chi_{\omega\omega}.
\end{align}
Therefore, we can investigate the validity of the EoS of spin fluids by comparing the invariants \eqref{eq:invariants} against the pseudo-gauge-independent expressions in \cref{eq:invariants-current}.

\paragraph{Ambiguity of spin equation of state}.---%
\Cref{eq:invariants} suggests that there is a pseudo-gauge ambiguity in the spin equation of state. Indeed, given a set of pseudo-gauge parameters solving \cref{eq:thermo-constraints}, we can always shift these by solutions to the homogeneous part of the PDEs. Up to quartic order in spin, these shifts are given as
\begin{align}
    \gamma_1 
    &\to \gamma_1 + \Lambda_1
    + a^2 \big(\Lambda_2 + \df_T(T\Lambda_4)\big)
    + \omega^2\lb \Lambda_3 + \Lambda_4 \rb, \nn\\
    \gamma_2 
    &\to \gamma_2 - a^2 \Lambda_4, \qquad
    \gamma_7
    \to \gamma_7 
    + 2\Lambda_2
    - \df_T(T\Lambda_4), \nn\\
    %
    \gamma_8 
    &\to \gamma_8 
    + \Lambda_4, \qquad 
    \gamma_{13}
    \to \gamma_{13}
    + \Lambda_3
    - \Lambda_4 , \nn\\
    \lambda_1 
    &\to \lambda_1 + \frac{\dow}{\dow\mu}
    \lb
    \Lambda_1
    + a^2 \Lambda_2 
    + \omega^2 \Lambda_3 \rb, \nn\\
    \lambda_2 
    &\to \lambda_2 
    -a^2 \frac{\dow\Lambda_4}{\dow\mu}, \qquad 
    \lambda_5
    \to \lambda_5
    -\frac{\dow\Lambda_4}{\dow\mu},
    %
    \label{eq:coefficient-shifts-EOS}
\end{align}
and similarly for $\lambda_{1,2,5}^S$ with $\dow/\dow\mu$ replaced by $\dow/\dow T$. Here the $\Lambda_i$'s are only functions of $T$ and $\mu$. 
These induce a pseudo-gauge improvement of the EoS
\begin{align}\label{eq:pressure-redef}
    p 
    &\to p 
    + a^2T\df_T\Lambda_1
    + 2\omega^2\Lambda_1
    + a^4 \lb 2\Lambda_2 + T\df_T\Lambda_2 \rb \nn\\
    &\qquad
    + 2\omega^4 \Lambda_3
    + a^2\omega^2 \lb
    4 \Lambda_2 + 4 \Lambda_3 
    + T \df_T\Lambda_3  \rb \nn\\
    &\qquad 
    + 2(a\cdot\omega)^2 
    \lb \Lambda_4 - \Lambda_2 - \Lambda_3\rb
    + \cO(\mu_{\mu\nu}^6).
\end{align}
One can verify that $\cI_2$ and $\cI_4$ in \cref{eq:invariants} are invariant under these improvements. Two representations of the EoS are only equivalent if their invariants match.


For conformal spin fluids, the EoS needs to preserve the structure in \cref{eq:conformal}, which forces all the $\Lambda_i$ parameters in \cref{eq:pressure-redef} to vanish. We have explicitly verified that the EoS of a conformal spin fluid can be identified unambiguously up to order $\cO(\mu_{\mu\nu}^{10})$, which is strongly indicative that it holds non-perturbatively.

\paragraph{Dirac fermions}.---Let us illustrate our procedure using the example of free Dirac fermions~\cite{Palermo:2021hlf, Palermo:2023ews, Ambrus:2014uqa, Ambrus:2019cvr, Ambrus:2019ayb, Buzzegoli:2017cqy, Becattini:2025oyi}. The respective currents in \cref{eq:general-currents} are given in \cref{app:micro-examples-fermion}. The constraint equations \eqref{eq:thermo-constraints} for the massless case admit a solution for pseudo-gauge parameters
\begin{align}
    \gamma_1
    &= \frac{1}{144} \left(T^2 + \frac{3 \mu ^2}{\pi^2}\right)
    + \frac{7a^2}{1440}
    - \frac{35 \omega^2}{1152\pi ^2}, \nn\\
    \gamma_2 
    &= \frac{-1}{72} \left(T^2+ \frac{3 \mu ^2}{\pi^2}\right)
    - \frac{11a^2}{5760 \pi^2}
    + \frac{7\omega^2}{480\pi^2}, \nn\\
    \gamma_7
    &= \frac{-1}{128 \pi ^2}, \quad 
    \gamma_8 = \frac{-91}{1920 \pi^2}, \quad 
    \gamma_{13} = \frac{17}{1920 \pi ^2}, \nn\\
    \lambda_1 
    &= \frac{-\mu}{8 \pi ^2}, \quad 
    \lambda_2 = \frac{-\mu}{12 \pi ^2}, \quad 
    \lambda_5 = 0.
    \label{eq:fermion-PG-params}
\end{align}
This casts the constitutive relations into the ideal spin fluid form, with the thermodynamic variables
\begin{gather}
    p 
    = \frac{7 \pi^2 T^4}{180}
    + \frac{T^2\mu^2}{6}
    + \frac{\mu^4}{12\pi ^2}
    + \frac{\omega^2}{36} \left(T^2 + \frac{3 \mu ^2}{\pi ^2}\right)
    - \frac{7\omega^4}{320 \pi ^2}, \nn\\
    n
    = 
    \frac{T^2\mu}{3} + \frac{\mu^3}{3\pi^2}
    + \frac{\mu\omega^2}{6\pi^2}, \qquad
    s = \frac{7\pi^2T^3}{45} 
    + \frac{T\mu^2}{3}
    + \frac{T\omega^2}{18}, \nn\\
    \chi_{\omega\omega}
    = \frac{T^2}{18} + \frac{\mu ^2}{6\pi^2}
    - \frac{7\omega^2}{80 \pi ^2},~~
    \chi_{aa},\chi_{a\omega} = 0,~~
    \epsilon = 3p.
    \label{eq:thermo-Dirac}
\end{gather}
Since the theory of massless free fermions is conformal, we have fixed the ambiguity \eqref{eq:pressure-redef} by requiring $p$ to take the conformal form in \cref{eq:conformal}. 
One can verify that our results satisfy local thermodynamic relations.
We can also evaluate the invariants \eqref{eq:invariants-current} for massless fermions
\begin{align}
    \cI_{2} 
    &=
    -\frac{T^2}{18} - \frac{\mu ^2}{6\pi^2}, \qquad
    \cI_{4} 
    =
    - \frac{7}{40 \pi^2},
    \label{eq:invariants-Dirac}
\end{align}
which agree with \cref{eq:invariants} evaluated on \cref{eq:thermo-Dirac}.

Our results qualitatively differ from~\cite{Becattini:2025oyi}. The thermodynamic quantities derived in \cite{Becattini:2025oyi} were shown to violate standard thermodynamic relations. Furthermore, the invariants evaluated on their EoS also differ from \cref{eq:invariants-Dirac}.

For massive free fermions, the currents have only been worked out to quadratic order in spin~\cite{Buzzegoli:2017cqy}. Since this theory is not conformal, we can only construct the respective EoS up to the $\Lambda_1$ ambiguity in \cref{eq:pressure-redef}. We find
\begin{align}
    p 
    &= \frac{m^4}{\pi^2}\!\sum_{n=1}^\infty(-)^{n+1}
    \lb \frac{K_2(x_n)}{x_n^2/2}
    - \frac{a^2}{T^2} \frac{K_3(x_n)}{24x_n/n^2}
    \rb \cosh\lb \frac{n\mu}{T} \rb \nn\\
    &\qquad 
    +  T\df_T\Lambda_1a^2
    + 2\Lambda_1\omega^2
    + \cO(\mu_{\mu\nu}^4),
    \label{eq:thermo-dirac-massive}
\end{align}
where $x_n = nm/T$, while $K_{2}$, $K_3$ are modified Bessel's functions. See \cref{app:micro-examples-fermion} for details. We can obtain other thermodynamic quantities from \cref{eq:thermo-dirac-massive}. The quadratic pseudo-gauge invariant is given as
\begin{align}
    \cI_2 = \frac{-m^4}{12\pi^2T^2}\!\sum_{n=1}^\infty(-)^{n+1}
    n^2
    \frac{K_3(x_n)}{x_n}
    \cosh\lb \frac{n\mu}{T} \rb.
    \label{eq:invariant-dirac-massive}
\end{align}

We have repeated this exercise for free scalar fields. The answers can be found in \cref{app:micro-examples-scalar}.

\paragraph{Discussion.}---%
In this work we have identified a family of thermodynamic pseudo-gauges, defined by \cref{eq:thermo-constraints}, where the densities and pressure appearing in the constitutive relations align with their thermodynamic values. We have also derived novel pseudo-gauge-independent thermodynamic relations \eqref{eq:thermo-relations} satisfied by equilibrium conserved currents.
We have used these results to obtain the spin-dependent thermodynamic variables and EoS for free Dirac fermions and scalar fields, thus solving the apparent contradiction between microscopic calculations and the coarse-grained spin hydrodynamics perspective.

We showed that, generically, measuring the equilibrium conserved currents only allows us to extract the thermodynamic variables up to certain pseudo-gauge ambiguities; see \cref{eq:pressure-redef}. Interestingly, for conformal spin fluids, this ambiguity is completely fixed by symmetry. It would be interesting to understand whether there are other physical principles that could constrain the EoS ambiguity in the non-conformal case. In this direction, it would be relevant to explore possible constraints arising from stability and causality in ideal spin fluids in an arbitrary pseudo-gauge.

Our derivation of the thermodynamic relations, the EoS pseudo-gauge ambiguity, and the pseudo-gauge-invariants relied on a perturbative expansion in spin. While this was sufficient to match with the microscopic results for free Dirac fermions and scalar fields, it would be interesting to explore whether one could derive a closed-form non-perturbative generalisation of these results.

Finally, it would be interesting to explore whether an action-based approach to spin hydrodynamics, along the work of~\cite{Grozdanov:2013dba, Harder:2015nxa, Crossley:2015evo, Haehl:2015uoc, Jensen:2017kzi}
and potentially involving approximate symmetries \cite{Armas:2021vku, Armas:2023tyx}, 
could help us isolate a preferred thermodynamic pseudo-gauge by, for instance, identifying preferred notions of stress tensor and spin current. We plan to address this in the future. 




\vspace{.5em}

\acknowledgements 

\paragraph{Acknowledgements.}---%
JA is partly funded by the Dutch Institute for Emergent Phenomena (DIEP) cluster at the University of Amsterdam via the DIEP
programme Foundations and Applications of Emergence
(FAEME) and the national NWA consortium Emergence At All Scales (EAAS). AJ is supported by the ERC Consolidator Grant GeoChaos (101169611). 

This work is partially funded by the European Union. Views and opinions expressed are however those of the authors only and do not necessarily reflect those of the European Union or the European Research Council Executive Agency. Neither the European Union nor the granting authority can be held responsible for them. For the purpose of open access, the authors have applied a CC BY public copyright licence to any Author Accepted Manuscript (AAM) version arising from this submission.

\bibliography{mySpires_ajain,extra}

\clearpage
\appendix

\setcounter{secnumdepth}{2}

\begin{widetext}

\section{Comments on spin thermodynamics}
\label{app:thermo}

In this appendix, we provide the derivation of spin thermodynamic relations in \cref{eq:spin-thermo}. Let us start with a simple example to understand the origin of the $\pi_\mu\df u^\mu$ term. Consider a term in pressure, $p\sim a^2$. It follows that
\begin{align}
    \df p \sim 2a^\mu \df a_\mu 
    = 2a^\mu \df \lb \mu_{\mu\nu} u^\nu \rb
    = 2a^{[\mu} u^{\nu]} \df \mu_{\mu\nu} 
    + 2a^\mu \mu_{\mu\nu} \df u^\nu ,
\end{align}
which leads to
\begin{align}
    \rho^{\mu\nu} \sim 4a^{[\mu} u^{\nu]}, \qquad 
    \pi^\mu \sim 2 a^\nu \mu_{\nu\rho}\Delta^\rho_\mu.
\end{align}
This illustrates the necessity of the $\pi_\mu\df u^\mu$ term in thermodynamics. Note that $u_\mu \df u^\mu = 0$ due to normalisation.

More generally, the pressure $p$ is an arbitrary function of the thermodynamic scalars: $T$, $\mu$, $a^2$, $\omega^2$, and $a\cdot\omega$. We can parametrise its differential as
\begin{align}
    \df p
    &= \frac{\dow p}{\dow T} \df T 
    + \frac{\dow p}{\dow\mu} \df\mu 
    + \frac{\dow p}{\dow a^2} \df a^2
    + \frac{\dow p}{\dow\omega^2} \df\omega^2
    + \frac{\dow p}{\dow(a\cdot\omega)} \df(a\cdot\omega) \nn\\
    &= \frac{\dow p}{\dow T} \df T 
    + \frac{\dow p}{\dow\mu} \df\mu 
    + \lb 2\frac{\dow p}{\dow a^2} a^\mu
    + \frac{\dow p}{\dow(a\cdot\omega)} \omega^\mu
    \rb \df a_\mu
    + \lb 2\frac{\dow p}{\dow\omega^2} \omega_\mu
    + \frac{\dow p}{\dow(a\cdot\omega)} a_\mu 
    \rb \df\omega^\mu \nn\\
    &= \frac{\dow p}{\dow T} \df T 
    + \frac{\dow p}{\dow\mu} \df\mu 
    + \lB  \lb 2\frac{\dow p}{\dow a^2} a^\mu
    + \frac{\dow p}{\dow(a\cdot\omega)} \omega^\mu
    \rb u^\nu 
    + \lb 2\frac{\dow p}{\dow\omega^2} \omega_\rho
    + \frac{\dow p}{\dow(a\cdot\omega)} a_\rho
    \rb \half \epsilon^{\mu\nu\sigma\rho} u_\sigma \rB
    \df \mu_{\mu\nu}
    \nn\\
    &\qquad 
    + \lB \lb 2\frac{\dow p}{\dow a^2} a^\nu
    + \frac{\dow p}{\dow(a\cdot\omega)} \omega^\nu
    \rb \mu_{\nu\lambda}
    + \lb 2\frac{\dow p}{\dow\omega^2} \omega^\nu
    + \frac{\dow p}{\dow(a\cdot\omega)} a^\nu 
    \rb \half \epsilon_{\lambda\nu\rho\sigma} \mu^{\rho\sigma} \rB
    \Delta^\lambda_\mu \df u^\mu.
\end{align}
Further using the identities
\begin{align}
    a^\nu \mu_{\nu\rho}\Delta^\rho_\mu 
    = - \ell_\mu , \qquad 
    \omega^\nu \mu_{\nu\rho}\Delta^\rho_\mu 
    = 0,\qquad
    \half \omega^\nu \epsilon_{\lambda\nu\rho\sigma} \mu^{\rho\sigma}\Delta^\lambda_\mu
    = -\ell_\mu, \qquad 
    \half a^\nu \epsilon_{\lambda\nu\rho\sigma} \mu^{\rho\sigma}
    \Delta^\lambda_\mu
    = 0,
\end{align}
this becomes
\begin{align}
    \df p
    &= \frac{\dow p}{\dow T} \df T 
    + \frac{\dow p}{\dow\mu} \df\mu 
    + \half \Big(  2\lb \chi_{aa} a^\mu
    + \chi_{a\omega}  \omega^\mu
    \rb u^\nu 
    + \lb \chi_{\omega\omega} \omega_\rho
    + \chi_{a\omega}  a_\rho
    \rb \epsilon^{\mu\nu\sigma\rho} u_\sigma \Big)
    \df \mu_{\mu\nu}
    - \lb \chi_{aa} + \chi_{\omega\omega} \rb \ell_\mu \df u^\mu,
    \label{eq:thermo-derivation}
\end{align}
where we have used
\begin{align}
    s = \frac{\dow p}{\dow T} , \qquad
    n = \frac{\dow p}{\dow\mu} , \qquad
    \chi_{aa} = 2\frac{\dow p}{\dow a^2}, \qquad 
    \chi_{\omega\omega} = 2\frac{\dow p}{\dow\omega^2}, \qquad     \chi_{a\omega} = \frac{\dow p}{\dow(a\cdot\omega)}.
\end{align}
\Cref{eq:thermo-derivation} is precisely the thermodynamic relation in \cref{eq:spin-thermo}, with $\rho^{\mu\nu}$ and $\pi_\mu$ given in \cref{eq:density-decomposition}.

We can also argue the form of the thermodynamic relation in \cref{eq:spin-thermo} via Lorentz symmetry. A Lorentz transformation $\Lambda_{\mu\nu}$ (satisfying $\Lambda^{\mu}_{~\rho} \Lambda_\mu{}^\nu = \delta_\rho^\nu$) acts on spin density  and chemical potential as
\begin{align}
    \rho^{\mu\nu}
    \to \Lambda^{\mu}_{~\rho}\Lambda^{\nu}_{~\sigma}
    \rho^{\rho\sigma}, \qquad 
    \mu_{\mu\nu}
    \to \Lambda_{\mu}^{~\rho}\Lambda_{\nu}^{~\sigma}
    \mu_{\rho\sigma}.
\end{align}
Note that $p$ is a scalar and hence is invariant under Lorentz transformations. This means that the differential $\df p$ should not depend on $\df\Lambda_{\mu\nu}$ after a Lorentz transformation. On the right-hand side, $s\df T$ and $n\df\mu$ are indeed invariant, but the spin term transforms as
\begin{align}
    \half \rho^{\mu\nu}\df \mu_{\mu\nu}
    \to \half \rho^{\mu\nu}\df \mu_{\mu\nu}
    + \rho_\mu{}^{\sigma} \mu_{\nu\sigma} 
    \lb \Lambda^{\rho\mu}
    \df \Lambda_{\rho}^{~\nu} \rb.
\end{align}
Note that the differential in the parenthesis is antisymmetric in $(\mu\leftrightarrow\nu)$. Therefore, the additional term does not vanish for $\rho_{[\mu}{}^{\sigma} \mu_{\nu]\sigma} \neq 0$. We can counter this with the $\pi_\mu \df u^\mu$ term. Using
\begin{align}
    u^\mu \to \Lambda^\mu_{~\rho} u^\rho, \qquad 
    \pi_\mu \to \Lambda_\mu^{~\rho} \pi_\rho,
\end{align}
we find
\begin{align}
    \pi_\mu\df u^\mu 
    \to \pi_\mu\df u^\mu 
    - u_\mu \pi_\nu
    \lb \Lambda^{\rho\mu} \df \Lambda_\rho^{~\nu} \rb.
\end{align}
Therefore, the right-hand side of the thermodynamic relation in \cref{eq:spin-thermo} is also invariant under Lorentz transformations, provided that we take
\begin{align}
    u_{[\mu} \pi_{\nu]}
    = \rho_{[\mu}{}^{\sigma} \mu_{\nu]\sigma} .
\end{align}
One may check that this rule is identically true for $\rho^{\mu\nu}$ and $\pi_\mu$ given in \cref{eq:density-decomposition}.

As an aside, note that we can also write the same thermodynamic relations in terms of the equilibrium thermodynamic variables: $T_0 = 1/\sqrt{-\beta_0^2}$, $u_0^\mu = T_0\beta_0^\mu$, $\mu_0 = T_0\nu_0$, and $\mu_{0}^{\mu\nu} = T_0\nu_{0}^{\mu\nu}$. To wit
\begin{align}
    \df p
    &= \hat s \df T_0
    + \frac{1}{2} \hat\rho^{\mu\nu} \df \mu_{0\mu\nu}
    + \hat\pi_\mu \df u^\mu_0 
    + \hat n \df\mu_0,  \nn\\
    \hat\epsilon
    &= T_0 \hat s + \half \mu_{0\mu\nu}
    \hat\rho^{\mu\nu} 
    + \mu_0 \hat n
    - p,
\end{align}
where we have defined the thermodynamic densities in the inertial frame of  $u^\mu_0$, given as
\begin{align}
    \hat\epsilon
    &= \epsilon
    + \mu_{0\mu\nu}x^{\nu} 
    \Gamma \Big( \pi^{\mu} + (\epsilon+p) u^{\mu} \Big), \nn\\
    \hat\rho^{\mu\nu}
    &= \Gamma \lb \rho^{\mu\nu} 
    + 2\pi^{[\mu} x^{\nu]} 
    + 2(\epsilon+p)u^{[\mu} x^{\nu]} 
    \rb, \nn\\
    \hat\pi_\mu 
    &= 
    \Gamma \Big( \pi^\nu  + \lb \epsilon + p \rb u^\nu \Big)
    \Delta_{0\mu\nu}, \nn\\
    \hat n 
    &= \Gamma n, \qquad 
    \hat s = \Gamma s,
\end{align}
with the Lorentz factor
\begin{align}
    \Gamma = \frac{T}{T_0} 
    = \sqrt{\frac{1}{1
    - 2u_{0}^\mu \mu_{0\mu\nu}x^\nu
    - \mu_0^{\mu\rho}\mu_{0\mu\nu} x_\rho x^\nu
    }}.
\end{align}
Note that the original unhatted thermodynamic densities are defined in the local rest frame of the fluid velocity $u^\mu$, which is non-inertial for spinning/boosted states.

\section{Useful identities in thermodynamic equilibrium}
\label{app:identities}

In this appendix, we collect some useful identities for spin fluids in thermodynamic equilibrium for later calculations. Firstly, note that the thermodynamic derivatives in the basis of scalars $\{T,\alpha^2,w^2,\alpha\cdot w,\nu\}$ are related to the ones in the basis $\{T,a^2,\omega^2,a\cdot\omega,\mu\}$ as
\begin{align}
    \df_T
    &= \frac{\dow}{\dow T} 
    + \frac{2a^2}{T}\frac{\dow}{\dow a^2} 
    + \frac{2\omega^2}{T}\frac{\dow}{\dow\omega^2}  
    + \frac{2(a\cdot\omega)}{T}\frac{\dow}{\dow(a\cdot\omega)}  
    + \frac{\mu}{T} \frac{\dow}{\dow\mu}
    , \nn\\ 
    \df_{\alpha^2}
    &= T^2\frac{\dow}{\dow a^2} , \qquad
    \df_{w^2}
    = T^2\frac{\dow}{\dow\omega^2} , \qquad
    \df_{(\alpha\cdot w)}
    = T^2\frac{\dow}{\dow (a\cdot\omega)} , \qquad
    \df_{\nu}
    = T\frac{\dow}{\dow\mu}, \nn\\
    \df_A 
    &\equiv 2 \df_{\alpha^2} + 2\df_{w^2}.
\end{align}
In these appendices, we will primarily use the non-scaled variables to avoid confusion, but will make use of $\df_T$ and $\df_A$ thermodynamic derivatives.

The spin chemical potential can be decomposed in terms of $a_\mu$ and $\omega_\mu$ as
\begin{align}
    \mu_{\mu\nu} 
    = 2u_{[\mu}a_{\nu]}
    + \bar\epsilon_{\mu\nu\rho}\omega^\rho,
\end{align}
where we have defined the spatial Levi-Civita symbol $\bar\epsilon^{\mu\nu\rho} = u_\lambda\epsilon^{\lambda\mu\nu\rho}$. We also note that
\begin{align}
    \ell^2 = a^2\omega^2 - (a\cdot\omega)^2.
\end{align}
The derivatives of various hydrodynamic variables in equilibrium are given by
\begin{align}
    \dow_{\mu}\frac{u_\nu}{T}
    = - \frac{\mu_{\mu\nu}}{T}, \quad 
    \dow_\mu\frac{\mu_{\rho\sigma}}{T}
    =\dow_\mu\frac{\mu}{T}
    = 0.
\end{align}
This can be used to derive
\begin{align}
    \frac{1}{T}\dow_\lambda T
    &= - a_\lambda, \nn\\
    \dow_\lambda u_\mu 
    &= - a_\mu u_\lambda 
    + \bar\epsilon_{\mu\lambda\rho}\omega^\rho, \nn\\
    \dow_\lambda a_\mu 
    &= - a_\mu a_\lambda
    - u_\mu u_\lambda a^2
    - \Delta_{\mu\lambda}\omega^2
    + \omega_\mu \omega_\lambda
    - 2u_{(\mu}\ell_{\lambda)}
    , \nn\\
    \dow_\lambda \omega_\mu
    &= 
    - 2\omega_\mu a_\lambda
    + \eta_{\mu\lambda} (a\cdot\omega), \nn\\
    \dow_\lambda \ell_\mu
    &= 
    - 3\ell_\mu a_\lambda
    - 2u_{(\mu} \bar\epsilon_{\lambda)\nu\rho} \ell^\nu \omega^\rho
    - \bar\epsilon_{\mu\lambda\rho} \omega^\rho \omega^2
    - \bar\epsilon_{\mu\lambda\nu} a^\nu (a\cdot\omega).
\end{align}
We also find
\begin{align}
    \frac{1}{2T}\dow_\mu \frac{a^2}{T^2}
    = \frac{1}{2T}\dow_\mu\frac{\omega^2}{T^2}
    &= - \omega^2a_\mu + (a\cdot\omega)\omega_\mu, \qquad
    \dow_\mu \frac{a\cdot\omega}{T^2}
    = 0.
\end{align}
In general, the derivative of an arbitrary function $f$ in equilibrium is given as
\begin{align}
    \dow_\mu f
    &=
    - a_\mu T\df_T f 
    - \lb \omega^2a_\mu - (a\cdot\omega) \omega_\mu  \rb 
    \frac{\df_A f}{T^2}.
\end{align}
In particular, note that $\ell^\mu \dow_\mu f = u^\mu\dow_\mu f = 0$.
Double derivatives of $f$ can similarly be expanded according to
\begin{align}
    \dow_\mu \dow_\nu f 
    &= 
    \bigg( a^2 T\df_T f 
    + \ell^2 \frac{\df_Af}{T^2}
    \bigg) u^\mu u^\nu 
    + \bigg( \omega^2 T\df_T f
    + \lb \omega^4 + a^2 \omega^2 - \ell^2 \rb \frac{\df_Af}{T^2}
    \bigg) \Delta^{\mu\nu} 
    \nn\\
    &\qquad
    + \bigg(
    2T \df_T f
    + 3\omega^2\frac{\df_A f}{T^2}
    + T^2 \df^2_T f
    + 2\omega^2 \frac{\df_T\df_A f}{T}
    + \omega^4 \frac{\df_A^2 f}{T^4}
    \bigg) a^\mu a^\nu \nn\\
    &\qquad
    - \bigg( 
    T\df_T f
    + \omega^2\frac{\df_A f}{T^2}
    - (a\cdot\omega)^2 \frac{\df_A^2 f}{T^4}
    \bigg) \omega^\mu \omega^\nu
    - 2(a\cdot\omega)
    \bigg( 
    \frac{2 \df_A f}{T^2}
    + \frac{\df_T\df_A f }{T}
    + \omega^2 \frac{\df_A^2 f}{T^4} \bigg) 
    a^{(\nu}\omega^{\mu)} \nn\\
    &\qquad 
    + 2 \bigg( T\df_T f
    + \omega^2 \frac{\df_A f}{T^2}
    \bigg)u^{(\mu} \ell^{\nu)}, \nn\\
    \dow^2 f 
    &= 
    \lb a^2 + 2\omega^2\rb T\df_T f 
    + a^2 T^2 \df^2_T f 
    + 2\omega^2\lb a^2 + \omega^2  \rb \frac{\df_Af }{T^2}
    + 2\ell^2 \frac{\df_T\df_A f}{T}
    + \ell^2 \omega^2 \frac{\df_A^2 f}{T^4}.
\end{align}
It is also possible to compute higher derivatives following the same steps if needed.

Requiring that the general Belinfante energy-momentum tensor $T^{\mu\nu}_\rmB$ given in \cref{eq:general-currents} is conserved in equilibrium leads to the following 4 identities among the components
\begin{align}\label{eq:EM-conservation-identity}
    \lb a^2 + 3\omega^2
    + a^2T\df_T
    + \frac{\ell^2}{T^2}\df_A\rb \cX_{aa}
    - (a\cdot\omega) \lb 2 - T\df_T \rb \mathcal{X}_{a \omega }
    + 2 \omega^2\cQ_{\ell}
    &= \cE 
    + \lb 
    1 - T\df_T
    - \frac{\omega ^2}{T^2}\df_A \rb \mathcal{P} , \nn\\
    - \cX_{aa}
    - \lb 1 - T\df_T \rb \cX_{\omega \omega }
    + \frac{1}{a\cdot\omega}
    \lb 
    2a^2+2\omega^2
    + a^2 T\df_T
    + \frac{\ell^2}{T^2} \df_A
    \rb \cX_{a\omega}
    - 2\cQ_\ell
    &= \frac{1}{T^2}\df_A\cP, \nn\\
    2\cQ_a
    - \bigg( 
    3a^2 + 4\omega^2 
    + a^2 T\df_T 
    + \frac{\ell^2}{T^2}\df_A \bigg) \cX_{a\ell}
    + (a\cdot\omega) \lb 1 - T\df_T \rb \cX_{\omega\ell}
    &= 0, \nn\\
    \lb 
    2\omega^2
    - a^2
    + a^2 T\df_T
    + \frac{\ell^2}{T^2}\df_A  \rb\cQ_{a}
    - (a\cdot\omega) \lb 3 - T\df_T \rb \cQ_\omega
    &= 0.
\end{align}
These identities are pseudo-gauge-independent, i.e. they apply in any pseudo-gauge. Also, note that the first two identities completely fix the $\cE$ and $\cQ_\ell$ components in terms of the others. The third identity completely fixes $\cQ_a$, while the fourth identity fixes $\cQ_\omega$ up to a term $\sim T^3f(\nu,\alpha^2,w^2,\alpha\cdot w)$.
Similarly, there is one identity arising from the conservation of charge current
\begin{align}\label{eq:charge-conservation-identity}
    \lb 2\omega^2 + a^2 T\df_T + \frac{\ell^2}{T^2}\df_A \rb\cN_a
    - (a\cdot \omega)\lb 2 - T\df_T \rb \cN_\omega
    &= 0~,
\end{align}
and a similar one for the entropy current
\begin{align}\label{eq:entropy-conservation-identity}
    \lb 2\omega^2 + a^2 T\df_T + \frac{\ell^2}{T^2}\df_A \rb\cS_a
    - (a\cdot \omega)\lb 2 - T\df_T \rb \cS_\omega
    &= 0.
\end{align}
These can be used to fix the $\cN_\omega$ and $\cS_\omega$ components up to terms $\sim T^2f(\nu,\alpha^2,w^2,\alpha\cdot w)$.





\section{Independent tensor structures for pseudo-gauge ambiguity}
\label{app:counting}

In this appendix we classify the tensor structures that can appear in the pseudo-gauge ambiguity tensors discussed in the main text. Since the vectors $u^\mu$, $a^\mu$, $\omega^\mu$, and $\ell^\mu$ form a complete spacetime basis, tensors can simply be constructed by performing a full decomposition along the basis vectors. Therefore, the number of independent tensor structures for a particular ambiguity tensor is equal to the number of its independent components.
Rank-2 antisymmetric tensors $M_J^{\mu\nu}$ and $M_S^{\mu\nu}$ have 6 independent tensor structures each. Rank-3 tensor $M_T^{\mu\nu\rho}$, which is antisymmetric in its last two indices, has 24 independent tensor structures. Lastly, rank-4 tensor $M_\Sigma^{\mu\nu\rho\sigma}$, which is antisymmetric in its first-two and last-two indices, has 36 independent structures. 

It is straightforward to write these structures out explicitly. However, we wish to organise tensor structures according to the maximum powers of spin chemical potential $\mu_{\mu\nu}$ they contain. We will call this the ``level'' of a tensor structure. To this end, we need to work in a basis that explicitly involves the fluid projection tensor $\Delta^{\mu\nu}$ and the transverse spin chemical potential $\bar\mu^{\mu\nu}\equiv \Delta^{\mu\rho}\Delta^{\nu\sigma}\mu_{\rho\sigma}$. Clearly, constructing tensor structures using all of $u^\mu$, $a^\mu$, $\omega^\mu$, $\ell^\mu$, $\Delta^{\mu\nu}$, and $\bar\mu^{\mu\nu}$ will inevitably lead to overcounting, which we will need to account for in the discussion below. It will be useful to note the following identities
\begin{align}
    \ell^2 \Delta^{\mu\nu}
    &= \ell^\mu \ell^\nu 
    + \omega^2 a^\mu a^\nu 
    + a^2 \omega^\mu \omega^\nu 
    - 2(a\cdot\omega) a^{(\mu}\omega^{\nu)}, \nn\\
    \ell^2 \bar\mu^{\mu\nu}
    &= -2\Big( \omega^2 a^{[\mu} - (a\cdot\omega)\omega^{[\mu} \Big) \ell^{\nu]}.
\end{align}
We have summarised the results in \cref{tab:rank-2-structures,tab:rank-3-structures,tab:rank-4-structures}. Note that we are only interested in parity-preserving tensors. The level-counting may be different if we allow for parity-violating structures.

\begin{table}[h]
    \renewcommand*{\arraystretch}{1.7}
    \centering
    \begin{tabular}{|c|c|c|c|}
    \hline
    \textbf{Level 1} &
    \textbf{Level 2} &
    \textbf{Level 3} &
    \textbf{Level 4} \\
    \hline
    \begin{minipage}[t]{10em}
        \vspace{-1.8em}
        \[
        \begin{aligned}
            \fJ_{1}^{\mu\nu} &\equiv 2u^{[\mu} a^{\nu]} \\ 
            \fJ_{2}^{\mu\nu} &\equiv \bar\mu^{\mu\nu}
        \end{aligned}
        \vspace{.5em}
        \]
    \end{minipage} &
    \begin{minipage}[t]{10em}
        \vspace{-1.8em}
        \[
        \begin{aligned}
            \fJ_{3}^{\mu\nu} \equiv 2u^{[\mu} \ell^{\nu]}
        \end{aligned}
        \]
    \end{minipage} &
    \begin{minipage}[t]{12em}
        \vspace{-1.8em}
        \[
        \begin{aligned}
            \fJ_{4}^{\mu\nu} &\equiv 2(a\cdot\omega)u^{[\mu} \omega^{\nu]} \\
            \fJ_{5}^{\mu\nu} &\equiv 2a^{[\mu} \ell^{\nu]}.
        \end{aligned}
        \vspace{.5em}
        \]
    \end{minipage} & 
    \begin{minipage}[t]{12em}
        \vspace{-1.8em}
        \[
        \begin{aligned}
            \fJ_{6}^{\mu\nu} \equiv 2(a\cdot\omega)a^{[\mu} \omega^{\nu]}
        \end{aligned}
        \]
    \end{minipage} \\
    \hline 
    \end{tabular}
    \caption{A basis of 6 independent rank-2 antisymmetric tensors.}
    \label{tab:rank-2-structures}
\end{table}

\begin{table}[h]
    \renewcommand*{\arraystretch}{1.7}
    \centering
    \begin{tabular}{|c|c|c|}
    \hline
    \textbf{Level 0} &
    \textbf{Level 1} &
    \textbf{Level 2} \\
    \hline
    \begin{minipage}[t]{10em}
        \vspace{-1.8em}
        \[
        \begin{aligned}
            \fT_{1}^{\mu\nu\rho} &\equiv 2\Delta^{\mu[\nu} u^{\rho]}
        \end{aligned}
        \]
    \end{minipage} & 
    \begin{minipage}[t]{10em}
        \vspace{-1.8em}
        \[
        \begin{aligned}
            \fT_{2}^{\mu\nu\rho} &\equiv 2u^\mu u^{[\nu} a^{\rho]} \\
            \fT_{3}^{\mu\nu\rho} &\equiv u^\mu \bar\mu^{\nu\rho} \\
            \fT_{4}^{\mu\nu\rho} &\equiv 2\bar\mu^{\mu[\nu} u^{\rho]} \\
            \fT_{5}^{\mu\nu\rho} &\equiv 2\Delta^{\mu[\nu} a^{\rho]}
        \end{aligned}
        \]
    \end{minipage} &
    \begin{minipage}[t]{10em}
        \vspace{-1.8em}
        \[
        \begin{aligned}
            \fT_{6}^{\mu\nu\rho} &\equiv 2u^\mu u^{[\nu} \ell^{\rho]} \\
            \fT_{7}^{\mu\nu\rho} &\equiv 2a^{\mu} a^{[\nu} u^{\rho]} \\
            \fT_{8}^{\mu\nu\rho} &\equiv 2\omega^{\mu} \omega^{[\nu} u^{\rho]} \\
            \fT_{9}^{\mu\nu\rho} &\equiv 2\Delta^{\mu[\nu} \ell^{\rho]} \\
            \fT_{10}^{\mu\nu\rho} &\equiv 2\bar\mu^{\mu[\nu} a^{\rho]} \\
            \fT_{11}^{\mu\nu\rho} &\equiv a^\mu\bar\mu^{\nu\rho}
        \end{aligned}
        \vspace{.5em}
        \]
    \end{minipage} \\
    \hline\hline
    \textbf{Level 3} &
    \textbf{Level 4} &
    \textbf{Level 5} \\
    \hline 
    \begin{minipage}[t]{12em}
        \vspace{-1.8em}
        \[
        \begin{aligned}
            \fT_{12}^{\mu\nu\rho} &\equiv 2(a\cdot\omega)u^{\mu} u^{[\nu} \omega^{\rho]} \\ 
            \fT_{13}^{\mu\nu\rho} &\equiv 2u^{\mu}a^{[\nu} \ell^{\rho]} \\
            \fT^{\mu\nu\rho}_{14} &\equiv 2a^\mu \ell^{[\nu} u^{\rho]} \\
            \fT^{\mu\nu\rho}_{15} &\equiv 2\ell^\mu a^{[\nu} u^{\rho]} \\ 
            \fT_{16}^{\mu\nu\rho} &\equiv 2(a\cdot\omega)\Delta^{\mu[\nu} \omega^{\rho]} \\
            \fT^{\mu\nu\rho}_{17} &\equiv 2\omega^\mu a^{[\nu} \omega^{\rho]}
        \end{aligned}
        \vspace{.5em}
        \]
    \end{minipage} & 
    \begin{minipage}[t]{12em}
        \vspace{-1.8em}
        \[
        \begin{aligned}
            \fT_{18}^{\mu\nu\rho} &\equiv 2(a\cdot\omega)u^{\mu}a^{[\nu} \omega^{\rho]} \\
            \fT_{19}^{\mu\nu\rho} &\equiv 2(a\cdot\omega)a^{\mu} \omega^{[\nu} u^{\rho]} \\
            \fT_{20}^{\mu\nu\rho} &\equiv 2(a\cdot\omega)\omega^{\mu} a^{[\nu} u^{\rho]} \\
            \fT^{\mu\nu\rho}_{21} &\equiv 2a^\mu a^{[\nu}\ell^{\rho]} \\
            \fT^{\mu\nu\rho}_{22} &\equiv 2\omega^\mu \omega^{[\nu}\ell^{\rho]}, \qquad
        \end{aligned}
        \]
    \end{minipage} &
    \begin{minipage}[t]{16em}
        \vspace{-1.8em}
        \[
        \begin{aligned}
            \fT^{\mu\nu\rho}_{23} &\equiv 2(a\cdot\omega)
            \Big(\omega^\mu \ell^{[\nu} 
            + \ell^\mu \omega^{[\nu}\Big) u^{\rho]} \\
            \fT^{\mu\nu\rho}_{24} &\equiv 2(a\cdot\omega)a^\mu a^{[\nu} \omega^{\rho]}
        \end{aligned}
        \]
    \end{minipage} \\
    \hline 
    \end{tabular}
    \caption{A basis of 24 independent rank-3 tensors antisymmetric in their last two indices.}
    \label{tab:rank-3-structures}
\end{table}

\begin{table}[h]
    \renewcommand*{\arraystretch}{1.7}
    \centering
    \begin{tabular}{|c|c|}
    \hline
    \textbf{Level 0} &
    \textbf{Level 1} \\
    \hline
    \begin{minipage}[t]{22em}
        \vspace{-1.8em}
        \[
        \begin{aligned}
            \fS_{1}^{\mu\nu\rho\sigma} &\equiv 4 u^{[\mu} \Delta^{\nu][\rho}u^{\sigma]} \\
            \fS_{2}^{\mu\nu\rho\sigma} &\equiv 2\Delta^{\mu[\rho}\Delta^{\sigma]\nu}.
        \end{aligned}
        \]
    \end{minipage} & 
    \begin{minipage}[t]{22em}
        \vspace{-1.8em}
        \[
        \begin{aligned}
            \fS_{3}^{\mu\nu\rho\sigma} &\equiv 4u^{[\mu}\bar\mu^{\nu][\rho} u^{\sigma]} \\
            \fS_{4,5}^{\mu\nu\rho\sigma} 
            &\equiv 4u^{[\mu}\Delta^{\nu][\rho} a^{\sigma]}
            \pm 4a^{[\mu}\Delta^{\nu][\rho} u^{\sigma]} \\
            \fS_{6}^{\mu\nu\rho\sigma} &\equiv 
            2\Delta^{\rho[\mu}\bar\mu^{\nu]\sigma}
            - 2\Delta^{\sigma[\mu}\bar\mu^{\nu]\rho}
        \end{aligned}
        \vspace{.5em}
        \]
    \end{minipage}  \\
    \hline\hline
    \textbf{Level 2} &
    \textbf{Level 3} \\ 
    \hline 
    \begin{minipage}[t]{12em}
        \vspace{-1.8em}
        \[
        \begin{aligned}
            \fS_{7}^{\mu\nu\rho\sigma} &\equiv 4u^{[\mu}a^{\nu]} a^{[\rho} u^{\sigma]} \\
            \fS_{8}^{\mu\nu\rho\sigma} &\equiv 4u^{[\mu}\omega^{\nu]}\omega^{[\rho} u^{\sigma]} \\
            \fS_{9,10}^{\mu\nu\rho\sigma} &\equiv 
            4u^{[\mu}\Delta^{\nu][\rho} \ell^{\sigma]}
            \pm 4\ell^{[\mu}\Delta^{\nu][\rho} u^{\sigma]} \\
            \fS_{11,12}^{\mu\nu\rho\sigma}  &\equiv 
            4u^{[\mu}\bar\mu^{\nu][\rho} a^{\sigma]}
            \pm 4a^{[\mu}\bar\mu^{\nu][\rho} u^{\sigma]} \\
            \fS_{13,14}^{\mu\nu\rho\sigma}  &\equiv 2u^{[\mu}a^{\nu]}\bar\mu^{\rho\sigma}
            \pm 2\bar\mu^{\mu\nu}u^{[\rho}a^{\sigma]} \\
            \fS_{15}^{\mu\nu\rho\sigma}  &\equiv 
            4a^{[\mu}\Delta^{\nu][\rho} a^{\sigma]} \\
            \fS_{16}^{\mu\nu\rho\sigma}  &\equiv 
            4\omega^{[\mu}\Delta^{\nu][\rho}\omega^{\sigma]}
            \vspace{.5em}
        \end{aligned}
        \vspace{.5em}
        \]
    \end{minipage} &
    \begin{minipage}[t]{22em}
        \vspace{-1.8em}
        \[
        \begin{aligned}
            \fG^{\mu\nu\rho\sigma}_{17,18} &\equiv 
            4u^{[\mu} a^{\nu]} \ell^{[\rho} u^{\sigma]}
            \pm 4u^{[\mu} \ell^{\nu]} a^{[\rho} u^{\sigma]} \\
            \fG_{19,20}^{\mu\nu\rho\sigma} &\equiv 
            4(a\cdot\omega)
            \lb u^{[\mu}\Delta^{\nu][\rho} \omega^{\sigma]}
            \pm \omega^{[\mu}\Delta^{\nu][\rho} u^{\sigma]}
            \rb \\
            \fS^{\mu\nu\rho\sigma}_{21,22} &\equiv 
           4u^{[\mu}\omega^{\nu]} \omega^{[\rho} a^{\sigma]}
            \pm 4a^{[\mu}\omega^{\nu]} \omega^{[\rho} u^{\sigma]} \\
            \fS_{23,24}^{\mu\nu\rho\sigma} &\equiv 
            4\ell^{[\mu}\Delta^{\nu][\rho} a^{\sigma]}
            \pm 4a^{[\mu}\Delta^{\nu][\rho} \ell^{\sigma]}
        \end{aligned}
        \vspace{.5em}
        \]
    \end{minipage}  \\
    \hline\hline
    \textbf{Level 4} &
    \textbf{Level 5} \\
    \hline 
    \begin{minipage}[t]{22em}
        \vspace{-1.8em}
        \[
        \begin{aligned}
            \fS_{25,26}^{\mu\nu\rho\sigma} &\equiv 
            4(a\cdot\omega) \lb u^{[\mu} a^{\nu]} \omega^{[\rho} u^{\sigma]}
            \pm u^{[\mu} \omega^{\nu]} a^{[\rho} u^{\sigma]} \rb \\
            \fS^{\mu\nu\rho\sigma}_{27,28} &\equiv 
            4u^{[\mu}a^{\nu]} a^{[\rho} \ell^{\sigma]}
            \pm 4\ell^{[\mu}a^{\nu]} a^{[\rho} u^{\sigma]} \\
            \fS^{\mu\nu\rho\sigma}_{29,30} &\equiv 
            4u^{[\mu}\omega^{\nu]} \omega^{[\rho} \ell^{\sigma]}
            \pm 4\ell^{[\mu}\omega^{\nu]} \omega^{[\rho} u^{\sigma]} \\
            \fS_{31,32}^{\mu\nu\rho\sigma} &\equiv 
            4(a\cdot\omega)\lb a^{[\mu}\Delta^{\nu][\rho} \omega^{\sigma]}
            \pm 4\omega^{[\mu}\Delta^{\nu][\rho}a^{\sigma]}\rb
        \end{aligned}
        \vspace{.5em}
        \]
    \end{minipage} &
    \begin{minipage}[t]{22em}
        \vspace{-1.8em}
        \[
        \begin{aligned}
            \fS^{\mu\nu\rho\sigma}_{33} &\equiv 4(a\cdot\omega)
            u^{[\mu}\Big(\omega^{\nu]} \ell^{[\rho} 
            + \ell^{\nu]} \omega^{[\rho}\Big) u^{\sigma]} \\
            \fS^{\mu\nu\rho\sigma}_{34,35} &\equiv 
            4(a\cdot\omega)
            \lb u^{[\mu}a^{\nu]} a^{[\rho} \omega^{\sigma]}
            \pm \omega^{[\mu}a^{\nu]} a^{[\rho} u^{\sigma]} \rb \\
            \fS_{36}^{\mu\nu\rho\sigma} &\equiv 
            4(a\cdot\omega)
            \lb \ell^{[\mu}\Delta^{\nu][\rho} \omega^{\sigma]}
            + 4\omega^{[\mu}\Delta^{\nu][\rho} \ell^{\sigma]} \rb
        \end{aligned}
        \]
    \end{minipage} \\
    \hline 
    \end{tabular}
    \caption{A basis of 36 independent rank-4 tensors antisymmetric in their first-two and last-two indices.}
    \label{tab:rank-4-structures}
\end{table}

\section{Details of improvement terms for charge current}
\label{app:PG-current}
The most general improvement tensor $M_J^{\mu\nu}$ for charge current may be expressed as
\begin{equation}
M^{\sigma\mu}_J = \sum_{i=1}^6 \lambda_i\, \fJ_i^{\sigma\mu},   
\end{equation}
where $\lambda_i$'s are arbitrary functions of the thermodynamic scalars $T$, $\mu$, $a^2$, $\omega^2$, and $a\cdot\omega$. We wish to compute its contribution to the equilibrium charge current as given in \cref{eq:pseudo-gauge}. Before we start, one may immediately see using the derivative identities that the contribution from the $\lambda_3$ term identically drops out. Using the parametrisation
\begin{equation}
    \partial_\sigma M^{\sigma\mu}_J
    = \delta\cN\, u^\mu
    + \delta\cN_a\,a^\mu 
    + \delta\cN_\omega\,\omega^\mu 
    + \delta\cN_\ell\,\ell^\mu,   
\end{equation}
for the remaining terms we find
\begin{align}\label{eq:U1-results}
    \delta\cN
    &= 
    \bigg( \lb a^2+2\omega^2\rb
    + a^2T\df_T
    + \frac{\ell^2}{T^2}\df_A \bigg) \lambda_1
    - 2\omega^2\lambda_2 
    + (a\cdot\omega)^2 \df_T(T\lambda_4)
    + 2\ell^2\lambda_5
    , \nn\\
    \delta\cN_\ell
    &= 
    \df_T(T\lambda_2)
    + \frac{\omega^2}{T^2}\df_A\lambda_2
    - \lambda_5 \lb 3a^2 + 2\omega^2\rb
    - a^2 T\df_T\lambda_5 
    - \frac{\ell^2}{T^2} \df_A \lambda_5, \nn\\
    \delta\cN_a
    &= (a\cdot\omega)^2\frac{1}{T} \df_T(T^2\lambda_6), \nn\\
    \delta\cN_\omega
    &= -2(a\cdot\omega)\omega^2\lambda_6 
    - (a\cdot\omega)a^2 \frac{1}{T^3}\df_T(T^4\lambda_6)
    - (a\cdot\omega)\frac{\ell^2}{T^2}\df_A \lambda_6.
\end{align}
We can verify that $\delta\cN_a$ and $\delta\cN_\omega$ satisfy the charge conservation identity in \cref{eq:charge-conservation-identity}.

Let us make a few interesting observations. Four coefficients $\lambda_{1,2,4,5}$ together only contribute to two components of the current $\delta\cN$ and $\delta\cN_\ell$. Therefore, it is plausible that there is some redundancy in these coefficients that does not affect $J^\mu$ at all. In general, finding such redundancies requires solving complicated partial differential equations. However, by brute force, we find the following redundancies
\begin{align}
    \lambda_{1} 
    &\to \lambda_1 
    + \frac{c^{(\lambda)}_1}{T}
    + \frac{c^{(\lambda)}_2}{T^5}\omega^2, \nn\\
    \lambda_2
    &\to \lambda_2 + \frac{c^{(\lambda)}_1}{T}
    - \frac{c^{(\lambda)}_2}{T^5} (a^2 - \omega^2)
    + \frac{T^2\alpha^2}{2} \df_T f_1^{(\lambda)}\,
    , \nn\\
    \lambda_4 
    &\to \lambda_4 + \frac{f_1^{(\lambda)}}{T}, \nn\\
    \lambda_5
    &\to \lambda_5 -\frac{c^{(\lambda)}_2}{T^5}
    + \half \df_Tf_1^{(\lambda)}~.
\end{align}
Here $c^{(\lambda)}$'s are functions of only $\nu$, $w^2-\alpha^2$, and $\alpha\cdot w$, while $f_1^{(\lambda)}$ is only a function of $T$, $\nu$, $w^2-\alpha^2$, and $\alpha\cdot w$. These are essentially some of the ``integration constants'' that would arise after solving the constraint equations.
In particular, we can use $f_1^{(\lambda)}$ to set $\lambda_4$ to zero at the leading order in spin.








\section{Details of improvement terms for Belinfante energy-momentum tensor}
\label{app:PG-stress}

We can repeat the exercise of the previous appendix to find improvements to the Belinfante energy-momentum tensor.
The general improvement tensor $M_\Sigma^{\sigma\mu\nu\rho}$ may be expressed as
\begin{equation}
    M^{\sigma\mu\nu\rho}_\Sigma = \sum_{i=1}^{36} \gamma_i\, \fS_i^{\sigma\mu\nu\rho},   
\end{equation}
where $\gamma_i$'s are functions of the thermodynamic scalars $T$, $\mu$, $a^2$, $\omega^2$, and $a\cdot\omega$. Its contribution can be decomposed as 
\begin{align}
    - \dow_\rho\dow_\sigma M^{\sigma(\mu\nu)\rho}_\Sigma
    &= \delta\cE\,u^\mu u^\nu 
    + \delta\cP\,\Delta^{\mu\nu}
    + \delta\cX_{aa}\, a^\mu a^\nu
    + \delta\cX_{\omega\omega}\,\omega^\mu \omega^\nu 
    + 2\delta\cX_{a\omega}\,a^{(\mu} \omega^{\nu)}
    + 2\delta\cQ_{\ell}\,u^{(\mu} \ell^{\nu)} \nn\\
    &\qquad 
    + 2\delta\cQ_{a} \,u^{(\mu} a^{\nu)}
    + 2\delta\cQ_\omega \,u^{(\mu} \omega^{\nu)}
    + 2\delta \cX_{a\ell} \,\alpha^{(\mu} \ell^{\nu)}
    + 2\delta \cX_{\omega\ell} \,\omega^{(\mu} \ell^{\nu)}
    .
\end{align}
Note that only the part of $M_\Sigma^{\rho\mu\nu\sigma}$ that is symmetrised in $(\mu\nu)$ and $(\rho\sigma)$ contribute to this expression. This means that the following 15 coefficients trivially do not contribute to the Belinfante tensor:
\begin{align}
    \gamma_3, \quad 
    \gamma_5, \quad 
    \gamma_6, \quad 
    \gamma_{10}, \quad 
    \gamma_{11}, \quad 
    \gamma_{14}, \quad 
    \gamma_{18}, \quad 
    \gamma_{20}, \quad 
    \gamma_{22}, \quad 
    \gamma_{24}, \quad 
    \gamma_{26}, \quad 
    \gamma_{28}, \quad 
    \gamma_{30}, \quad 
    \gamma_{32}, \quad 
    \gamma_{35}.
\end{align}
We find that the combination $\fS_{12}^{\rho\mu\nu\sigma}+\fS_{13}^{\rho\mu\nu\sigma}$ is also identically zero when symmetrised over $(\mu\nu)$ and $(\rho\sigma)$. Therefore, the respective coefficients only appear in the combination $\gamma_{12}-\gamma_{13}$ and, without loss of generality, we can drop
\begin{align}
    \gamma_{12}.
\end{align}

This leaves us with 20 coefficients parametrising improvements to the Belinfante tensor. Out of these, 12 coefficients: $\gamma_1$, $\gamma_2$, $\gamma_7$, $\gamma_8$, $\gamma_9$, $\gamma_{13}$, $\gamma_{15}$, $\gamma_{16}$, $\gamma_{25}$, $\gamma_{27}$, $\gamma_{29}$, and $\gamma_{31}$ only contribute to the following components
{\allowdisplaybreaks
\begin{subequations}
\begin{align}
    \delta\cE
    &= - \bigg( 
    6 \omega^2
    + 2(a^2+\omega^2) \lb T\df_T + \frac{\omega^2}{T^2}\df_A \rb
    + T^2a^2\df^2_T
    + \frac{\ell^2 }{T^2}
    \lb \df_A + 2T\df_T\df_A + \frac{\omega^2}{T^2}\df_A^2 \rb
    \bigg) \gamma_1 
    + 6\omega^2\gamma_2 \nn\\
    &\qquad 
    - \bigg(
    6(a^2 + \omega^2)^2
    + a^2 \Big( 
    (6a^2+4\omega^2) T\df_T
    + a^2 T^2\df_T^2
    \Big)
    \nn\\
    &\hspace{6em}
    + \ell^2
    \Big( 14+3 T\df_T
    + 7 \frac{a^2+\omega^2}{T^2} \df_A
    + 2 \frac{a^2}{T^2} T\df_T\df_A
    + \frac{\ell^2}{T^4}\df_A^2
    \Big)
    \bigg) \gamma_7 \nn\\
    &\qquad 
    - (a\cdot\omega)^2 
    \bigg(2 T\df_T + T^2\df_T^2\bigg)\gamma_8
    + 6 \bigg( 
    2 \left(\ell^2+\omega^4+a^2\omega^2\right)
    + \ell^2 T\df_T
    + \frac{\ell^2\omega^2}{T^2}\df_A
    \bigg)\gamma_9
    \nn\\
    &\qquad 
    - 2 \bigg(
    6\left(\ell^2+\omega^4+a^2 \omega^2\right) 
    + \left(\ell^2+2 a^2 \omega^2\right) T\df_T
    + 3\ell^2 \omega^2 \df_A
    \bigg)\gamma_{13} 
    - 6\ell^2\gamma_{15}\nn \\
    &\qquad
    - 2(a\cdot\omega)^2
    \bigg(
    2(5a^2+2\omega^2)
    + 2(4a^2+\omega^2)T\df_T
    + a^2T^2\df_T^2
    + \frac{\ell^2}{T^2}(2\df_A+T\df_T\df_A)
    \bigg) \gamma_{25}\nn \\
    &\qquad 
    + 6\bigg(
    2(3a^2+2\omega^2)\ell^2
    + a^2 \ell^2 T \df_T
    + \frac{\ell^4}{T^2}\df_A
    \bigg)\gamma_{27}
    , \\[0.5em]
    %
    %
    \delta\cP 
    &= - \bigg( 
    2\omega^2
    - Ta^2 \df_T
    - \frac{\ell^2}{T^2} \df_A 
    \bigg) \gamma_1 
    + \bigg( 
    2\omega^2
    + T\omega^2\df_T 
    + \frac{\omega^2}{T^2}(a^2+\omega^2) \df_A
    + T^2 a^2\df^2_T
    + \ell^2 
    \lb 
    \frac{2}{T}\df_A\df_T
    + \frac{1}{T^4}\omega^2
    \df_A^2 \rb
    \bigg)\gamma_2 \nn\\
    &\qquad 
    - 2\ell^2 \gamma_7 
    + 2\bigg(
    2\left(\ell^2+\omega^4+a^2 \omega^2\right)
    + \ell^2 T\df_T
    + \frac{\ell^2 \omega^2}{T^2}\df_A
    \bigg) \gamma_9
    \nn\\
    &\qquad 
    - 2 \bigg(
    2\left(\ell^2+\omega^4+a^2 \omega^2\right) 
    + a^2 \omega^2 T\df_T
    + \frac{\ell^2 \omega^2}{T^2} \df_A
    \bigg) \gamma_{13} \nn \\
     &\qquad
    - \bigg(
    2(a^2+\omega^2)^2
    + 10\ell^2
    + \lb 4a^4+2a^2\omega^2+3\ell^2\rb T\df_T
    + a^4T^2\df_T^2 \nn\\
    &\hspace{6em}
    + 5(a^2+\omega^2)\frac{\ell^2}{T^2}\df_A
    + \frac{2a^2 \ell^2}{T^2} T\df_T\df_A 
    + \frac{\ell^4}{T^4}\df_A^2 \bigg)\gamma_{15}
     \nn \\
    &\qquad
    - (a\cdot\omega)^2\lb 2T\df_T 
    + T^2\df_T^2\rb\gamma_{16} 
    + 2\ell^2\bigg(
    2(3a^2+2\omega^2)
    + a^2 T\df_T
    + \frac{\ell^2}{T^2}\df_A \bigg)\gamma_{27}\nn \\
    &\qquad
    - 2(a\cdot\omega)^2\bigg(
    2(4a^2+\omega^2)
    + (7a^2+\omega^2)T\df_T
    + a^2T^2\df_T^2
    + \frac{\ell^2}{T^2}(2\df_A+T\df_T\df_A)
    \bigg)\gamma_{31}
    ,\\[0.5em]
    %
    %
    \delta\cX_{aa} 
    &= - 2\bigg( T \df_T \gamma_1 + \frac{\omega^2}{T^2}\df_A\gamma_1
    \bigg) 
    - \bigg(
    \frac{\omega^2}{T^2}\df_A 
    + T^2 \df^2_T 
    + 2 \frac{\omega^2}{T} \df_T\df_A
    + \frac{\omega^4}{T^4}\df_A^2 
    \bigg) \gamma_2  \nn\\
    &\qquad 
    - \bigg(
    2\left(a^2+\omega^2\right)
    + a^2 T\df_T
    + \frac{\ell^2}{T^2}\df_A
    \bigg)\gamma_7\nn\\
    &\qquad  
    + \bigg(
    2(a^2+5\omega^2)
    + 4(a^2+\omega^2)T\df_T
    + a^2 T^2\df_T^2 \nn\\
    &\hspace{6em}
    + \lb 4\omega^4 + 4a^2\omega^2 + \ell^2\rb
    \frac{1}{T^2}\df_A
    + \frac{2a^2 \omega^2}{T}\df_T\df_A
    + \frac{\omega^2\ell^2}{T^4}\df_A^2
    \bigg)\gamma_{15} \nn \\
    &\qquad
    -2(a\cdot\omega)^2(T\df_T+2)\gamma_{25}
    +2(a\cdot\omega)^2(T\df_T+2)\gamma_{27}\nn\\
    &\qquad
    +2(a\cdot\omega)^2
    \bigg( 8 + 7T\df_T + T^2\df_T^2
    + 2\omega^2\df_A + \frac{\omega^2}{T}\df_T\df_A \bigg)\gamma_{31} 
    ,\\[0.5em]      
    %
    %
    \delta\cX_{\omega\omega}
    &= 4\gamma_1
    - \bigg( 
    4
    - T\df_T
    - \frac{\omega^2}{T^2}\df_A 
    + \frac{(\alpha\cdot w)^2}{T^4} \df_A^2
    \bigg) \gamma_2
    + 2a^2\gamma_7
    + \bigg( 
    2(2 a^2+\omega^2)
    + a^2 T\df_T
    + \frac{\ell^2}{T^2}\df_A
    \bigg)
    (\gamma_8 - 4\gamma_9 + 2\gamma_{13}) \nn\\
    &\qquad 
    - 2\bigg(
    (a^2+2\omega^2)
    + a^2 T\df_T
    - \lb a^2 \omega^2-2\ell^2\rb \frac{1}{T^2}\df_A
    \bigg)\gamma_{15}\nn \\
    &\qquad 
    - \bigg( 
    6 \left(2 a^2+\omega^2\right)
    + 2 \left(4 a^2+\omega^2\right)T\df_T
    + a^2 T^2\df_T^2
    + \lb 5 \ell^2+2\omega^4+4 a^2 \omega^2\rb \frac{1}{T^2}\df_A
    + \frac{2 a^2 \omega^2}{T} \df_T\df_A
    \bigg) \gamma_{16} \nn\\
    &\qquad 
    - 2\bigg( 
    2 \left(3a^4+\ell^2+a^2 \omega^2\right)
    + a^4 T\df_T
    + \frac{\ell^2 a^2}{T^2} \df_A
    \bigg)\gamma_{27}
    - 2\bigg( 
    2 \omega^2 \left(4 a^2+\omega^2\right)
    + \lb 2 a^2 \omega^2-\ell^2\rb T\df_T
    \bigg)\gamma_{29} \nn\\
    &\qquad 
    - 2(a\cdot\omega)^2 \bigg( 
    2 + T\df_T
    + 2 (3a^2+\omega^2) \frac{1}{T^2}\df_A
    + \frac{a^2}{T}\df_T\df_A
    + \frac{\ell^2}{T^4}\df_A^2
    \bigg) \gamma_{31}
    , \\[0.5em]
    %
    %
    \frac{\delta\cX_{a\omega}}{a\cdot\omega}
    &= \frac{1}{T^2}\df_A \gamma_1
    + \frac{1}{T^2}
    \bigg( 
    \df_A
    + T \df_T\df_A
    + \frac{\omega^2}{T^2} \df_A^2 \bigg) \gamma_2
    - \gamma_7
    - \df_T(T\gamma_8)
    + \frac{2}{T}\df_T(T^2\gamma_9)
    - 2\gamma_{13} \nn\\
    &\qquad 
    - \bigg(
    5 + T\df_T 
    + \lb 3 a^2+4 \omega^2\rb \frac{1}{T^2}\df_A
    + \frac{a^2}{T}\df_T\df_A 
    + \frac{\ell^2}{T^4} \df_A^2
    \bigg)\gamma_{15} \nn\\
    &\qquad 
    + \bigg(
    3 + 5T\df_T + T^2 \df_T^2
    + \frac{\omega^2}{T^2} \df_A
    + \frac{\omega^2}{T}\df_T\df_A 
    \bigg)\gamma_{16}
    + 2a^2 \gamma_{27}
    + 2\frac{\omega^2}{T^2}\df_T(T^3\gamma_{29})
    + \frac{2(a\cdot\omega)^2}{T^2}\df_A\gamma_{31}
    , \\[0.5em]
    %
    %
    \delta\cQ_\ell
    &= \bigg( T \df_T
    + \frac{\omega^2}{T^2} \df_A  \bigg) 
    \Big(2\gamma_1 - 3\gamma_2\Big)
    + \bigg(
    2 \left(a^2+\omega^2\right)
    + a^2 T\df_T
    + \frac{\ell^2}{T^2} \df_A
    \bigg) \gamma_7 \nn\\
    &\qquad 
    - \bigg(
    8 (a^2 + 2\omega^2)
    + \lb 7 a^2 + 4 \omega^2 \rb T\df_T
    + a^2 T^2\df_T^2
    + 4 \left(\ell^2+\omega^4+a^2 \omega^2\right) \frac{1}{T^2}\df_A
    + \frac{2 \ell^2}{T} \df_T\df_A
    + \frac{\ell^2 \omega^2}{T^4}\df_A^2
    \bigg)\gamma _9 \nn\\
    &\qquad 
    + \bigg(
    8( a^2 + 2\omega^2)
    + \lb 7 a^2 + 4\omega^2\rb T\df_T
    + a^2 T^2\df_T^2
    + 4 \left(\ell^2+\omega^4+a^2 \omega^2\right)\frac{1}{T^2} \df_A
    \nn\\
    &\hspace{6em}
    + \lb \ell^2 + a^2 \omega^2\rb \frac{1}{T}\df_T\df_A
    + \frac{\ell^2 \omega^2}{T^4} \df_A^2
    \bigg) \gamma_{13} \nn\\
    &\qquad 
    + 3\bigg( 
    2(a^2+\omega^2) 
    + a^2 T\df_T
    + \frac{\ell^2}{T^2} \df_A
    \bigg) \gamma_{15}
    + \frac{(a\cdot\omega)^2}{T} \df_T(T^2\gamma_{25}) \nn\\
    &\qquad 
    - \bigg( 
    2\lb 12 a^4+11 \ell^2+4 \omega^4+13 a^2 \omega^2 \rb
    + \lb 11 a^4+3 \ell^2+5 a^2 \omega^2\rb T\df_T
    + a^4 T^2\df_T^2 \nn\\
    &\hspace{6em}
    + 4\ell^2 \left(3 a^2+2 \omega^2\right) \frac{1}{T^2}\df_A
    + \frac{2a^2 \ell^2}{T}\df_T\df_A
    + \frac{\ell^4}{T^4} \df_A^2
    \bigg)\gamma_{27} \nn\\
    &\qquad 
    - (a\cdot\omega)^2 \bigg( 
    6 + 6T\df_T + T^2 \df_T^2
    \bigg)\gamma_{29}
    + 3(a\cdot\omega)^2 \bigg( 
    2 + T\df_T
    \bigg)\gamma_{31}.
\end{align}
\label{eq:EM-results-even}%
\end{subequations}
The remaining 8 coefficients: $\gamma_4$, $\gamma_{17}$, $\gamma_{19}$, $\gamma_{21}$, $\gamma_{23}$, $\gamma_{33}$, $\gamma_{34}$, and $\gamma_{36}$ only contribute to the following components
\begin{subequations}
\begin{align}
    \frac{\delta\cQ_a}{(a\cdot\omega)^2}
    &= - \frac{1}{T^2}\bigg( 
    \df_A
    - T\df_T\df_A
    \bigg)\gamma_4
    - \bigg(1+T\df_T\bigg)\gamma_{17} 
    + \bigg(
    5 + 7T\df_T + T^2\df_T^2
    + \frac{\omega^2}{T^2}\lb \df_A + T\df_T\df_A\rb
    \bigg)\gamma_{19} \nn \\
    &\qquad  
    - \bigg( 2 +4T\df_T + T^2\df_T^2\bigg)\gamma_{21} 
    - 3\bigg(1 +T\df_T\bigg)\gamma_{23}
    - \omega^2\bigg(3+T\df_T\bigg)\gamma_{33}\nn \\
    &\qquad  
    + \bigg(
    3(7a^2+3\omega^2)
    + (11a^2+3\omega^2)T\df_T
    + a^2T^2\df_T^2
    + \frac{3\ell^2}{T^2}\df_A
    + \frac{\ell^2}{T^2} T\df_T\df_A
    \bigg)\gamma_{34}
    - 3\omega^2\bigg(3+T\df_T\bigg)\gamma_{36}
    , \\[0.5em]
    %
    %
    \frac{\delta\cQ_\omega}{a\cdot\omega}
    &= -
    \bigg( 
    \lb a^2 + 2\omega^2 \rb \frac{1}{T^2} \df_A
    + \frac{a^2}{T} \df_T\df_A
    + \frac{\ell^2}{T^4}\df_A^2 \bigg)\gamma_4
    + \bigg(3a^2+2\omega^2
    + a^2T\df_T
    + \frac{\ell^2}{T^2}\df_A\bigg)\gamma_{17} \nn \\
    &\qquad  
    - \bigg(
    5\lb 3a^2+2\omega^2\rb
    + (9a^2+2\omega^2) T\df_T
    + a^2T^2\df_T^2 \nn\\
    &\hspace{10em}
    + \lb 3a^2\omega^2+2w^4+7\ell^2\rb \frac{1}{T^2}\df_A
    + \lb a^2\omega^2+\ell^2\rb \frac1T \df_T\df_A
    + \frac{\omega^2\ell^2}{T^4}\df_A^2
    \bigg)\gamma_{19} \nn \\
    &\qquad
    + \bigg(
    2\lb 3a^2+2\omega^2\rb
    + 2(3a^2+\omega^2)T\df_T
    + T^2a^2\df_T^2
    + \frac{2\ell^2}{T^2}\df_A
    + \frac{\ell^2}{T}\df_T\df_A
    \bigg)\gamma_{21}\nn \\
    &\qquad
    + 3\bigg((3a^2+2\omega^2)+a^2 T\df_T
    + \frac{\ell^2}{T^2}\df_A\bigg)\gamma_{23}
    + \bigg(
    5a^2\omega^2+2\omega^4+2\ell^2
    + a^2\omega^2T\df_T
    + \frac{\omega^2\ell^2}{T^2}\df_A\bigg)\gamma_{33}\nn \\
    &\qquad
    - \bigg(
    35a^4+29a^2\omega^2+6\omega^4+20\ell^2
    - (13a^4+5a^2 \omega^2+2\ell^2)T\df_T
    + a^4T^2\df_T^2 \nn\\
    &\hspace{10em}
    + 7(2a^2+\omega^2)\ell^2\frac{1}{T^2}\df_A
    + \frac{2a^2\ell^2}{T}\df_T\df_A
    + \frac{\ell^4}{T^4}\df_A^2 \bigg)\gamma_{34}\nn \\
    &\qquad
    + 3\bigg(
    5a^2\omega^2+2\ell^2+2\omega^4
    + a^2 \omega^2T\df_T
    + \frac{\omega^2\ell^2}{T^2}\df_A \bigg)\gamma_{36}
    \\[0.5em]
    %
    %
    \frac{\delta\cX_{a\ell}}{(a\cdot\omega)^2}
    &= \frac{1}{T^2}\df_T\lb T\df_A\gamma_{23}\rb
    - (3+T\df_T) (\gamma_{33}-\gamma_{34}) 
    + \bigg(
    18 + 10T\df_T + T^2\df_T^2
    + \frac{3\omega^2}{T^2}\df_A
    + \frac{\omega^2}{T}\df_T\df_A
    \bigg)\gamma_{36}
    , \\[0.5em]
    %
    %
    \frac{\delta\cX_{\omega\ell}}{a\cdot\omega}
    &= \frac{2}{T^2}\df_A\gamma_4
    - 2\gamma_{17}
    + 2\bigg(5+T\df_T
    +\frac{\omega^2}{T^2}\df_A\bigg)\gamma_{19}
    - 2\bigg(2+T\df_T\bigg)\gamma_{21}\nn \\
    &\qquad
    - \bigg(6
    + (5a^2+4\omega^2)\frac{1}{T^2}\df_A
    + \frac{a^2}{T}\df_T\df_A
    + \frac{\ell^2}{T^4}\df_A^2\bigg)\gamma_{23}
    + \bigg(7a^2+2\omega^2
    + a^2T\df_T
    + \frac{\ell^2}{T^2}\df_A
    \bigg)(\gamma_{33}+\gamma_{34}) \nn \\
    &\qquad
    - \bigg(
    42a^2
    + 30 \omega^2
    + \lb 14a^2 + 4\omega^2\rb T\df_T
    + a^2 T^2\df_T^2 \nn\\
    &\hspace{10em}
    + \lb 7 a^2 \omega^2 + 4\omega^4 + 8\ell^2\rb  \frac{1}{T^2} \df_A
    + \lb a^2\omega^2 + \ell^2 \rb \frac{1}{T} \df_T\df_A
    + \frac{\ell^2\omega^2}{T^4} \df_A^2
    \bigg) \gamma_{36}. 
\end{align}
\label{eq:EM-results-odd}%
\end{subequations}
}
One may check that these expressions satisfy the conservation identities in \cref{eq:EM-conservation-identity}.

Similar to our discussion in the U(1) case, we can identify redundancies in the coefficients that leave the Belinfante energy-momentum tensor invariant. A subset of these is given as
\begin{align}
    \gamma_1 
    &\to \gamma_1 
    + c_1^{(\gamma)}
    + a^2 \lb 
    \frac{3c_3^{(\gamma)}}{T^4}
    + f_2^{(\gamma)}
    - 2 T\df_Tf_1^{(\gamma)}
    - 2 T\df_Tf_3^{(\gamma)} 
    - T^2\df_T^2f_1^{(\gamma)}
    - \frac{1}{2}T^2\df_T^2f_2^{(\gamma)}
    - \frac{5}{2} T^2\df_T^2f_3^{(\gamma)}
    - \frac{1}{2}T^3\df_T^3f_3^{(\gamma)}
    \rb \nn\\
    &\qquad 
    + \omega^2 \lb 
    - \frac{c_3^{(\gamma)}+5c_4^{(\gamma)}}{2T^4}
    + f_2^{(\gamma)}
    - T\df_Tf_1^{(\gamma)}
    - \frac{1}{2}T\df_Tf_2^{(\gamma)} 
    - T\df_Tf_3^{(\gamma)}
    - \frac{1}{2}T^2\df_T^2f_3^{(\gamma)}
    \rb ,
    \nn\\
    \gamma_2 
    &\to \gamma_2 + c_1^{(\gamma)}
    + a^2 \lb 
    \frac{c_3^{(\gamma)}}{T^4}
    + T\df_T f_1^{(\gamma)}
    + \half T\df_T f_2^{(\gamma)}
    + T\df_T f_3^{(\gamma)}
    + \frac{1}{2} T^2\df_T^2f_3^{(\gamma)}
    \rb
    - \frac{c_4^{(\gamma)}}{T^4}\omega^2
    , \nn\\
    \gamma_7
    &\to \gamma_7 
    - \frac{3c_3^{(\gamma)}}{T^4}
    - f_2^{(\gamma)}
    + 2 T\df_Tf_1^{(\gamma)}
    + 2 T\df_Tf_3^{(\gamma)}
    + T^2\df_T^2f_1^{(\gamma)}
    + \frac{1}{2}T^2\df_T^2f_2^{(\gamma)}
    + \frac{5}{2}T^2\df_T^2f_3^{(\gamma)}
    + \frac{1}{2}T^3\df_T^3f_3^{(\gamma)}, \nn\\
    \gamma_8 
    &\to \gamma_8 
    - \frac{3c_3^{(\gamma)}-3c_4^{(\gamma)}}{2T^4}
    + 2f_1^{(\gamma)}
    + 2f_2^{(\gamma)}
    + 3f_3^{(\gamma)}
    - T\df_Tf_1^{(\gamma)}
    - \frac{1}{2}T\df_Tf_2^{(\gamma)}
    - \frac{1}{2}T^2\df_T^2f_3^{(\gamma)}, \nn\\
    \gamma_{9}
    &\to \gamma_{9} + f_1^{(\gamma)}, \nn\\
    \gamma_{13}
    &\to \gamma_{13}
    + \frac{3c_3^{(\gamma)}+3c_4^{(\gamma)}}{4T^4}
    + f_1^{(\gamma)} 
    - f_2^{(\gamma)}
    + T\df_Tf_1^{(\gamma)}
    + \frac{1}{2}T\df_Tf_2^{(\gamma)}
    + T\df_Tf_3^{(\gamma)}
    + \frac{1}{2}T^2\df_T^2f_3^{(\gamma)}, \nn\\
    \gamma_{15}
    &\to \gamma_{15} + f_2^{(\gamma)}, \nn\\
    \gamma_{16}
    &\to \gamma_{16} + f_3^{(\gamma)}.
\end{align}
Here $c^{(\gamma)}$'s are functions of only $\nu$, $w^2-\alpha^2$, and $\alpha\cdot w$, while $f^{(\gamma)}$'s are functions of $T$, $\nu$, $w^2-\alpha^2$, and $\alpha\cdot w$. We can use $f^{(\gamma)}_{1,2,3}$ to set $\gamma_{9,15,16}$ to zero at leading order in spin.

\section{Details of thermodynamic constraints and thermodynamic relations}
\label{app:details-thermo-constrains}

We wish to obtain the pseudo-gauge improvements that cast the equilibrium conserved currents in a generic pseudo-gauge in \cref{eq:general-currents} into ideal spin fluid constitutive relations given in \cref{eq:ideal-fluid,eq:ideal-belinfante}. Let us denote the transformed components of the conserved currents with a prime, e.g. $\cP' = \cP + \delta\cP$. Firstly, we get the constraints
\begin{equation}\label{eq:trivial-constraints}
    \cX_{a\ell}' 
    = \cX_{\omega\ell}' 
    = \cQ_a' 
    = \cQ_\omega'
    = \cN_{a}' 
    = \cN_\omega'
    = \cS_{a}' 
    = \cS_\omega' = 0.
\end{equation}
After substituting the pseudo-gauge improvements from \cref{eq:U1-results,eq:EM-results-odd}, these yield PDEs for 10 pseudo-gauge parameters: $\gamma_4$, $\gamma_{17}$, $\gamma_{19}$, $\gamma_{21}$, $\gamma_{23}$, $\gamma_{33}$, $\gamma_{34}$, $\gamma_{36}$, $\lambda_6$, and $\lambda^S_6$. In all the microscopic examples we discuss, the respective components of the currents are already zero. Therefore, we can trivially solve the constraints in \cref{eq:trivial-constraints} by setting the relevant pseudo-gauge parameters to zero. Of course, these parameters may be more non-trivial to solve for in generic microscopic theories. The remaining constraints are given as
\begin{gather}
    \cE'
    + 2 \omega^2\lb \cX'_{\omega\omega} -\cX'_{aa}\rb
    - a^2\df_T\lb T\cX'_{aa}\rb
    - \frac{\ell^2}{T^2} \df_A \cX'_{aa}
    = 
    \epsilon
    - (a\cdot\omega)\df_T (T\chi_{a\omega}),
    \nn\\
    2\cQ'_\ell
    - T\df_T\cX'_{\omega\omega}
    - \frac{\omega^2}{T^2}\df_A\cX'_{\omega\omega}
    =
    \frac{a\cdot\omega}{T^2}\df_A \chi_{a\omega}, \nn\\
    \cP'
    + \omega^2\cX'_{\omega\omega}
    = p
    - (a\cdot\omega)\chi_{a\omega}, \nn\\
    \cX'_{aa}
    = -\chi_{aa}, \qquad 
    \cX'_{\omega\omega}
    = \chi_{\omega\omega}, \qquad
    \cN' 
    = n, \qquad
    \cS' 
    = s, \qquad
    \cX_{a\omega}'
    = \cN_\ell'
    = \cS_\ell'
    = 0,
\end{gather}
which provide PDEs to solve for 20 pseudo-gauge parameters: $\gamma_1$, $\gamma_2$, $\gamma_7$, $\gamma_8$, $\gamma_9$, $\gamma_{13}$, $\gamma_{15}$, $\gamma_{16}$, $\gamma_{25}$, $\gamma_{27}$, $\gamma_{29}$, $\gamma_{31}$, $\lambda_1$, $\lambda_2$, $\lambda_4$, $\lambda_5$, $\lambda_1^S$, $\lambda_2^S$, $\lambda_4^S$, and $\lambda_5^S$.
Using the thermodynamic relations on the right-hand side, we find that two combinations of these reduce to the first two conservation identities in \cref{eq:EM-conservation-identity}. The third equation becomes the formula for EoS in \cref{eq:EOS-formula}, while the remaining combinations yield the thermodynamic constraints outlined in \cref{eq:thermo-constraints}.

Solving these PDEs in general is extremely hard. However, we can make some headway by working perturbatively in spin. Note that $\cX_{aa}$, $\cX_{\omega\omega}$, $\cQ_\ell$, $\cN_\ell$, and $\cS_\ell$ only begin contributing to the constitutive relations at $\cO(\mu_{\mu\nu}^2)$. On the other hand, parity symmetry requires that $\cX_{a\omega} \propto (a\cdot\omega)$ and hence only begins contributing at $\cO(\mu_{\mu\nu}^4)$. Therefore, at zeroth order in spin, the only non-trivial part of the constraints \eqref{eq:thermo-constraints} is
\begin{align}
    \lim_{\mu_{\mu\nu}\to0}\cN' - \frac{\dow\cP'}{\dow\mu}=0, \qquad 
    \lim_{\mu_{\mu\nu}\to0}\cS' - \frac{\dow\cP'}{\dow T}=0.
\end{align}
These quantities turn out to be independent of pseudo-gauge improvements and instead enforce thermodynamic relations on the original currents as discussed in the main text. At leading order in spin, i.e. at $\cO(\mu_{\mu\nu}^2)$, we get the following independent constraints 
\begin{gather}
    \cX'_{aa}
    + 2\frac{\dow\cF'}{\dow a^2}
    = \cO(\mu_{\mu\nu}^2),  \qquad
    \cX_{\omega\omega}'
    - 2\frac{\dow\cF'}{\dow\omega^2}
    = \cO(\mu_{\mu\nu}^2), \nn\\
    \frac{\dow}{\dow a^2}\lb \cN'
    - \frac{\dow\cF'}{\dow\mu} \rb
    = \cO(\mu_{\mu\nu}^2), \qquad
    \frac{\dow}{\dow \omega^2}\lb \cN'
    - \frac{\dow\cF'}{\dow\mu} \rb
    = \cO(\mu_{\mu\nu}^2), \qquad
    \cN_\ell = \cO(\mu_{\mu\nu}^2). \nn\\
    \frac{\dow}{\dow a^2}\lb \cS'
    - \frac{\dow\cF'}{\dow T} \rb
    = \cO(\mu_{\mu\nu}^2), \qquad
    \frac{\dow}{\dow \omega^2}\lb \cS'
    - \frac{\dow\cF'}{\dow T} \rb
    = \cO(\mu_{\mu\nu}^2), \qquad 
    \cS_\ell = \cO(\mu_{\mu\nu}^2).
    \label{eq:leading-order-constraints}
\end{gather}
Note that the constraint on $\cN'$ splits into two constraints for $\dow\cN'/\dow a^2$ and $\dow\cN'/\dow\omega^2$, and similarly for $\cS'$. At leading order the pseudo-gauge improvements in \cref{eq:EM-results-even,eq:U1-results} reduce to
\begin{align}
    \delta\cE
    &= - a^2\Big( 
    2 T\df_T\gamma_1 
    + T^2\df^2_T\gamma_1 
    \Big) 
    - \omega^2\Big( 6\gamma_1
    + 2T\df_T
    - 6\gamma_2\Big)
    + \cO(\mu_{\mu\nu}^4), \nn\\
    %
    %
    \delta\cP 
    &= a^2\Big( 
    T\df_T\gamma_1
    + T^2 \df^2_T\gamma_2
    \Big)
    - \omega^2\Big( 
    2\gamma_1
    - 2\gamma_2
    - T\df_T\gamma_2
    \Big)
    + \cO(\mu_{\mu\nu}^4), \nn\\
    %
    %
    \delta\cX_{aa} 
    &= - 2T \df_T \gamma_1
    - T^2 \df^2_T \gamma_2
    + \cO(\mu_{\mu\nu}^2), \nn\\   
    %
    %
    \delta\cX_{\omega\omega}
    &= 4\gamma_1
    - 4\gamma_2 + T\df_T\gamma_2
    + \cO(\mu_{\mu\nu}^2), \nn\\
    %
    %
    \delta\cX_{a\omega}
    &= \cO(\mu_{\mu\nu}^2), \nn\\
    %
    %
    \delta\cQ_\ell
    &= 2T \df_T \gamma_1
    - 3T\df_T\gamma_2
    + \cO(\mu_{\mu\nu}^2), \nn\\
    \delta\cN
    &= 
    a^2\Big( \lambda_1 + T\df_T\lambda_1 \Big) 
    + \omega^2\Big(2\lambda_1 -2\lambda_2 \Big)
    + \cO(\mu_{\mu\nu}^4)
    , \nn\\
    \delta\cN_\ell
    &= 
    \df_T(T\lambda_2)
    + \cO(\mu_{\mu\nu}^2), \nn\\
    \delta\cS
    &= 
    a^2\Big( \lambda^S_1 + T\df_T\lambda^S_1 \Big) 
    + \omega^2\Big(2\lambda^S_1 -2\lambda^S_2 \Big)
    + \cO(\mu_{\mu\nu}^4)
    , \nn\\
    \delta\cS_\ell
    &= 
    \df_T(T\lambda_2^S)
    + \cO(\mu_{\mu\nu}^2).
\end{align}
All pseudo-gauge parameters at this order can be viewed as only functions of $T$ and $\mu$. Further note that
\begin{align}
    \delta\cF
    = \delta\cP + \omega^2 \delta\cX_{\omega\omega}
    &= 
    a^2\Big( 
    T\df_T\gamma_1
    + T^2 \df^2_T\gamma_2
    \Big)
    + \omega^2\Big( 
    2\gamma_1 - 2\gamma_2 
    + 2T\df_T \gamma_2
    \Big) 
    + \cO(\mu_{\mu\nu}^4).
\end{align}
Substituting these back into \cref{eq:leading-order-constraints}, we find the PDEs
\begin{align}
    -T^2 \df^2_T\gamma_2
    &= \cX_{aa}
    + 2\frac{\dow\cF}{\dow a^2}
    + \cO(\mu_{\mu\nu}^2),  \nn\\
    3T\df_T \gamma_2
    &= \cX_{\omega\omega}
    - 2\frac{\dow\cF}{\dow\omega^2}
    + \cO(\mu_{\mu\nu}^2), \nn\\
    \df_T\lb T\lambda_1 - T\frac{\dow}{\dow\mu}\Big( 
    \gamma_1
    - \gamma_2
    + T\df_T\gamma_2
    \Big) \rb
    &= -
    \frac{\dow}{\dow a^2}\lb \cN
    - \frac{\dow\cF}{\dow\mu} \rb
    + \cO(\mu_{\mu\nu}^2), \nn\\
    \lambda_1 - \lambda_2
    - \frac{\dow}{\dow\mu}\Big( 
    \gamma_1 - \gamma_2 
    + T\df_T \gamma_2
    \Big) 
    &= -\frac{1}{2}
    \frac{\dow}{\dow \omega^2}\lb \cN
    - \frac{\dow\cF}{\dow\mu} \rb
    + \cO(\mu_{\mu\nu}^2), \nn\\
    \df_T(T\lambda_2) &= - \cN_\ell
    + \cO(\mu_{\mu\nu}^2), \nn\\
    \df_T\lb T\lambda^S_1 - T\frac{\dow}{\dow T}\Big( 
    \gamma_1
    - \gamma_2
    + T\df_T\gamma_2
    \Big) \rb
    &= -
    \frac{\dow}{\dow a^2}\lb \cS
    - \frac{\dow\cF}{\dow T} \rb
    + \cO(\mu_{\mu\nu}^2), \nn\\
    \lambda_1^S - \lambda_2^S
    - \frac{\dow}{\dow T}\Big( 
    \gamma_1 - \gamma_2 
    + T\df_T \gamma_2
    \Big) 
    &= -\frac{1}{2}
    \frac{\dow}{\dow \omega^2}\lb \cS
    - \frac{\dow\cF}{\dow T} \rb
    + \cO(\mu_{\mu\nu}^2), \nn\\
    \df_T(T\lambda_2^S) &= - \cS_\ell
    + \cO(\mu_{\mu\nu}^2).
\end{align}
We can eliminate $\gamma_2$ from the first two equations to yield the Belinfante tensor thermodynamic relation in \cref{eq:thermo-relations}. On the other hand, we can use the next three and the last three equations to derive the particle current and entropy current thermodynamic relation in \cref{eq:thermo-relations} respectively. Solving the remaining equations yield the solution
\begin{align}
    \gamma_1 
    &= \Lambda_1 + \gamma_2 - T\df_T\gamma_2, \nn\\
    \gamma_2
    &= c^{(\gamma)}_1 + \int \frac{\df T}{3T} \lb \cX_{\omega\omega}
    - 2\frac{\dow\cF}{\dow\omega^2} \rb, \nn\\
    \lambda_1 
    &= \frac{c^{(\lambda)}_1}{T} 
    + \frac{\dow\Lambda_1}{\dow\mu}
    - \frac{1}{2}
    \frac{\dow}{\dow \omega^2}\lb \cN
    - \frac{\dow\cF}{\dow\mu} \rb
    - \frac1T \int\df T\, \cN_\ell 
    + \cO(\mu_{\mu\nu}^2), \nn\\
    \lambda_2 &= \frac{c^{(\lambda)}_1}{T} 
    - \frac1T \int\df T\, \cN_\ell 
    + \cO(\mu_{\mu\nu}^2) , \nn\\
    \lambda_1^S
    &= \frac{c^{(S)}_1}{T} 
    + \frac{\dow\Lambda_1}{\dow T}
    - \frac{1}{2}
    \frac{\dow}{\dow \omega^2}\lb \cS
    - \frac{\dow\cF}{\dow T} \rb
    - \frac1T \int\df T\, \cS_\ell 
    + \cO(\mu_{\mu\nu}^2), \nn\\
    \lambda_2^S &= \frac{c^{(S)}_1}{T} 
    - \frac1T \int\df T\, \cS_\ell 
    + \cO(\mu_{\mu\nu}^2),
\end{align}
where $\Lambda_1$ is an arbitrary function of $T$ and $\mu$, while the $c$'s are only functions of $\mu/T$.
The $\df T$ integrals here are performed at constant $\mu/T$. Using \cref{eq:EOS-formula}, the EoS at this order is simply
\begin{align}
    p = \cF + T\df_T\Lambda_1a^2
    + 2\Lambda_1\omega^2
    + \cO(\mu_{\mu\nu}^4).
\end{align}
We have recovered the $\Lambda_1$ ambiguity of the EoS discussed in the main text. The sub-leading order computation can be found in the supplementary Mathematica notebook.

\section{Details of non-interacting Dirac fermions}
\label{app:micro-examples-fermion}

The coefficients appearing in the Belinfante energy-momentum tensor and charge current in \cref{eq:general-currents} for massless non-interacting Dirac fermions are given as~\cite{Palermo:2021hlf}
\begin{align}\label{eq:currents-Dirac}
    \cE
    &= 
    \frac{7\pi^4T^4 + 30 \pi^2 T^2 \mu^2 + 15 \mu^4}{60\pi^2}
    + \frac{\pi^2 T^2 + 3 \mu ^2}{24\pi^2} (a^2+3\omega^2)
    - \frac{17a^4}{960\pi^2}
    + \frac{\omega^4}{64\pi^2} 
    + \frac{23a^2\omega^2}{1440\pi^2}
    +\frac{11(a\cdot\omega)^2}{720\pi^2}, \nn\\
    \cP
    &= \frac13 \cE
    - \frac{2}{27 \pi ^2}
    \Big( a^2\omega^2 - (a\cdot\omega)^2 \Big)
    , \nn\\
    \cX_{aa}
    &= \frac{a^2}{9\pi^2}, \qquad 
    \cX_{\omega\omega}
    = \frac{\omega^2}{9\pi^2}, \qquad 
    \cX_{a\omega}
    = \frac{(a\cdot\omega)}{9\pi^2}, \qquad 
    \cX_{a\ell}
    = \cX_{\omega\ell} = 0,
    \nn\\
    \cQ_\ell
    &= \frac{\pi^2 T^2+3\mu^2}{18\pi^2} 
    + \frac{31a^2}{360\pi^2}
    + \frac{13\omega^2}{120\pi^2}, \qquad 
    \cQ_{a}
    = \cQ_{\omega} = 0,\nn\\
    \cN
    &= \frac{\mu}{\pi^2}
    \lb \frac{\pi^2 T^2 + \mu^2}{3} 
    + \frac{a^2+\omega^2}{4}
    \rb, \nn\\
    \cN_\ell
    &= \frac{\mu}{6\pi^2}, \qquad 
    \cN_{a}
    = \cN_{\omega} = 0.
\end{align}
Note that the authors in~\cite{Palermo:2021hlf, Becattini:2020qol, Buzzegoli:2017cqy, Becattini:2025oyi} are using the conventions $\eta_{\mu\nu}=\diag(1,-1,-1,-1)$ and $\epsilon^{0123}=1$, as opposed to our conventions of $\eta_{\mu\nu}=\diag(-1,1,1,1)$ and $\epsilon_{0123}=1$. We have presented the pseudo-gauge parameters needed to cast \cref{eq:currents-Dirac} into the thermodynamic form in \cref{eq:fermion-PG-params}. The ensuing equation of state is given in \cref{eq:thermo-Dirac}. More details on the derivation of these results can  be found in the supplementary Mathematica notebook.

For the massive case, the expressions for conserved currents have only been derived up to leading order in spin~\cite{Buzzegoli:2017cqy}:
\begin{align}
    \cE 
    &= \frac{2m^4}{\pi^2}\sum_{n=1}^\infty
    (-1)^{n+1}
    \cosh (\nu  n) \left(
    \lb 1 + \frac{n^2\omega^2}{8T^2} \rb \left(
    \frac{K_3\left(x_n\right)}{x_n}
    - \frac{K_2\left(x_n\right)}{x_n^2}\right) 
    + \frac{n^2a^2}{24T^2} \left(1+\frac{3}{x_n^2}\right) K_2\left(x_n\right)
    \right) +\cO(\mu_{\mu\nu}^4), \nn\\
    \cP
    &= \frac{2m^4}{\pi^2}\sum_{n=1}^\infty
    (-1)^{n+1}
    \cosh (\nu  n) \left(
    \left(1+\frac{n^2\omega^2}{8T^2}\right)
    \frac{K_2\left(x_n\right)}{x_n^2}
    + \frac{n^2a^2}{24T^2} \left(\frac{K_3\left(x_n\right)}{x_n}-\frac{3 K_2\left(x_n\right)}{x_n^2}\right)
    \right)  + \cO(\mu_{\mu\nu}^4), \nn\\
    \cX_{aa}
    &= \cX_{\omega\omega}
    = \cX_{a\omega}
    = \cX_{a\ell}
    = \cX_{\omega\ell} = \cO(\mu_{\mu\nu}^2), \nn\\
    \cQ_\ell
    &= \frac{2m^4}{\pi^2}\sum_{n=1}^\infty
    (-1)^{n+1}
    \cosh (\nu  n) \frac{-n^2}{24T^2}
    \frac{K_3(x_n)}{x_n}
    + \cO(\mu_{\mu\nu}^2), \nn\\
    \cN
    &= \frac{2m^3}{\pi^2}\sum_{n=1}^\infty
    (-1)^{n+1}
    \sinh (\nu  n) 
    \left(
    \left(1+\frac{n^2\omega^2}{8T^2}\right)
    \frac{K_2\left(x_n\right)}{x_n}
    + \frac{n^2a^2}{24T^2} \left(
    K_3\left(x_n\right)
    -\frac{K_2\left(x_n\right)}{x_n}\right)
    \right)  + \cO(\mu_{\mu\nu}^4), \nn\\
    \cN_\ell
    &= \frac{2m^3}{\pi^2}\sum_{n=1}^\infty
    (-1)^{n+1}
    \sinh (\nu  n) 
    \frac{n^2}{12T^2}
    \frac{K_2(x_n)}{x_n}
    + \cO(\mu_{\mu\nu}^2), \qquad 
    \cN_{a}
    = \cN_{\omega} = \cO(\mu_{\mu\nu}^2)~,
\end{align}
where $x_n = n m/T$ and $K_{\alpha}(x_n)$ are modified Bessel's functions. At this order, the pseudo-gauge improvement parameters to cast these into thermodynamic form are given as
\begin{align}
    \gamma_1 
    &= \Lambda_1
    + \frac{m^2}{12\pi^2}\sum_{n=1}^\infty
    (-1)^{n+1}
    \cosh (\nu  n) \left(K_0\left(x_n\right)
    +K_2\left(x_n\right)\right)
    + \cO(\mu_{\mu\nu}^2), \nn\\
    \gamma_2
    &= -\frac{m^2}{6\pi^2}\sum_{n=1}^\infty
    (-1)^{n+1} 
    \cosh (\nu  n) 
    \frac{K_1\left(x_n\right)}{x_n}
    + \cO(\mu_{\mu\nu}^2), \nn\\
    \lambda_1 
    &= \frac{\dow\Lambda_1}{\dow\mu}
    - \frac{m}{6\pi^2}\sum_{n=1}^\infty
    (-1)^{n+1}\sinh (\nu  n) 
    K_1\left(x_n\right)
    + \cO(\mu_{\mu\nu}^2), \nn\\
    \lambda_2 
    &= -\frac{m}{6\pi^2}\sum_{n=1}^\infty
    (-1)^{n+1} \sinh (\nu  n) 
    K_1\left(x_n\right)
    + \cO(\mu_{\mu\nu}^2),
\end{align}
for arbitrary function $\Lambda_1(T,\mu)$. The respective equation of state in given in \cref{eq:thermo-dirac-massive}, while the quadratic pseudo-gauge-invariant is given in \cref{eq:invariant-dirac-massive}.

\section{Details of  non-interacting scalar fields}
\label{app:micro-examples-scalar}

Similarly for massless non-interacting scalar fields, the components of the Belinfante energy-momentum tensor and charge current are given as~\cite{Becattini:2020qol}
\begin{align}\label{eq:massless-scalar-results}
    \cE
    &= 
    \frac{8\pi^4 T^4
    + 60 \pi ^2 T^2\mu^2- 15 \mu ^4
    }{120\pi^2}
    + \left(T^2 -\frac{3 \mu^2}{2\pi^2}\right)
    \lb \frac{1-4\xi}{6}  \omega^2
    + \frac{1-6\xi}{6} a^2\rb \nn\\
    &\qquad 
    - \frac{11-60\xi}{240 \pi ^2}a^4
    - \frac{1-4\xi}{24 \pi ^2}\omega ^4
    - \frac{61-270 \xi}{360\pi^2} a^2 \omega^2
    + \phi_1 (a\cdot\omega)^2, \nn\\
    \cP 
    &= \frac13 \cE 
    + \left(T^2-\frac{3\mu^2}{2\pi^2} \right)
    \lb \frac{10 \xi -1}{18} \omega^2
    - \frac{1-6 \xi}{6} a^2\rb
    + \frac{1-4\xi}{24\pi ^2}a^4
    - \frac{10 \xi-1}{72\pi ^2}\omega^4
    - \frac{90 \xi-7}{108\pi ^2} a^2 \omega^2
    + \phi_2 (a\cdot\omega)^2, \nn\\
    \cX_{aa}
    &= \left(T^2-\frac{3 \mu^2}{2 \pi ^2}\right)\frac{1-6\xi}{6} 
    - \frac{1-6 \xi}{24\pi ^2} a^2
    + \frac{42 \xi -1}{72\pi ^2}\omega^2, \nn\\
    \cX_{\omega\omega}
    &= - \left(T^2-\frac{3 \mu^2}{2 \pi ^2}\right)\frac{1-2\xi}{6}
    + \frac{11-6 \xi }{72 \pi ^2}a^2
    + \frac{1-2\xi}{24 \pi ^2}\omega^2, \nn\\
    \cX_{a\omega} 
    &= \phi_3 (a\cdot\omega), \qquad
    \cX_{a\ell}
    = \cX_{\omega\ell} = 0, \nn\\
    \cQ_\ell 
    &= - \left(T^2-\frac{3 \mu^2}{2\pi^2}\right) 
    \frac{(6 \xi +1)}{9}
    - \frac{7-30 \xi}{360 \pi^2} a^2
    + \frac{10 \xi -1}{120 \pi ^2}\omega^2, \qquad 
    \cQ_{a}
    = \cQ_{\omega} = 0, \nn\\
    \cN 
    &= \frac{\mu}{3}\left(T^2 - \frac{\mu^2}{2\pi^2}\right), \qquad 
    \cN_\ell 
    = \frac{\mu}{6\pi^2}, \qquad
    \cN_a = \cN_\omega = 0,
\end{align}
where $\xi$ is an arbitrary parameter arising from coupling to curved spacetime~\cite{Becattini:2020qol}. A few comments are in order. Massless scalar fields only admit equilibrium states for $\mu=0$. Hence, we can only read off the $\mu\to 0$ part of \cref{eq:massless-scalar-results} by comparing with the currents in equilibrium. We have reinstated the $\mu\neq 0$ part of the constitutive relations by comparing with the results for massive scalar fields in the $m\to 0$ limit~\cite{Buzzegoli:2017cqy}; see below. As such, \cref{eq:massless-scalar-results} should be interpreted in the hydrodynamic sense, valid for small perturbations near equilibrium.

Furthermore, the energy-momentum tensor in the original reference \cite{Becattini:2020qol} was only computed for states with either nonzero vorticity or acceleration, and is therefore missing contributions proportional to $a\cdot\omega$. We have remedied this by allowing $\cE$ and $\cP$ to contain arbitrary terms proportional to $(a\cdot\omega)^2$, and $\cX_{a\omega}$ to contain a linear term in $(a\cdot\omega)$, controlled by constant coefficients $\phi_{1,2,3}$. It turns out that two of these coefficients are fixed by requiring that the energy-momentum tensor in identically conserved in equilibrium, i.e.
\begin{align}
    \phi_2 = \frac{15 \xi -1}{27 \pi ^2}, \qquad 
    \phi_3 = -\frac{12 \xi +1}{36\pi ^2}.
\end{align}
The remaining coefficient $\phi_1$, as well as the coupling $\xi$, can be removed via pseudo-gauge improvements.

The parameters of pseudo-gauge improvements to cast \cref{eq:massless-scalar-results} into conformal thermodynamic form are given as
\begin{align}
    \gamma_1 
    &=
    \frac{1-6\xi}{36} \left(T^2-\frac{3 \mu ^2}{2 \pi ^2}\right) 
    + 
    a^2 \lb 
    \frac{96 \xi -19}{864 \pi ^2}
    + \frac{\phi_1}{6}\rb
    + \omega^2\lb \frac{300 \xi -77}{4320 \pi ^2}
    + \frac{\phi }{6} \rb, \nn\\
    \gamma_2 
    &=
    \frac{1-6\xi}{36} \left(T^2-\frac{3 \mu ^2}{2 \pi ^2}\right)
    + a^2 \lb \frac{19-60 \xi }{2160 \pi ^2}
    - \frac{\phi }{6}\rb
    + \frac{\omega^2 }{120 \pi ^2}, \nn\\
    \gamma_7 
    &= \frac{31-150 \xi }{2160 \pi ^2} -\frac{\phi }{6}, \qquad 
    \gamma_8
    = \frac{75 \xi -56}{1080 \pi ^2} + \frac{\phi }{6}, \qquad 
    \gamma_{13}
    = \frac{71-300 \xi }{4320 \pi ^2} -\frac{\phi }{6}.
\end{align}
The resultant EoS is given as
\begin{align}
    \boxed{
    p = \frac{\pi^4 T^4}{45\pi^2}
    + \frac{T^2\mu^2}{6}
    - \frac{\mu^4}{24\pi^2}
    - \left(T^2 - \frac{3 \mu ^2}{2\pi^2}\right)
    \frac{\omega^2}{18}
    - \frac{\omega^4}{80 \pi ^2}.}
\end{align}
Let us also compute the remaining thermodynamic observables
\begin{gather}
    \epsilon 
    = 3p, \qquad 
    n
    = \frac{T^2\mu}{3} - \frac{\mu^3}{6\pi ^2 }
    + \frac{\mu\omega^2}{6\pi^2}, \qquad 
    s = \frac{4\pi^2 T^3}{45}
    + \frac{T\mu^2}{3}
    - 
    \frac{T\omega^2}{9}, \nn\\
    \chi_{\omega\omega}
    = - \frac{T^2}{9} + \frac{\mu ^2}{6\pi^2}
    - \frac{\omega^2}{20 \pi ^2}, \qquad 
    \chi_{aa} = \chi_{a\omega} = 0.
\end{gather}
One can also compute the quadratic and quartic pseudo-gauge invariants
\begin{align}
    \cI_2 =  \frac{T^2}{9} - \frac{\mu ^2}{6\pi^2}, \qquad 
    \cI_4 =  
    - \frac{1}{10 \pi ^2}.
\end{align}

For massive complex scalar fields, the results for the conserved currents at leading order in spin are given as~\cite{Buzzegoli:2017cqy}
\begin{align}\label{eq:massless-scalar-results-massive}
    \cE
    &= 
    \frac{m^4}{\pi ^2}\sum_{n=1}^\infty
    \cosh (\nu  n) \Bigg(
    \frac{K_3\left(x_n\right)}{x_n}
    - \frac{K_2\left(x_n\right)}{x_n^2}
    + \frac{n^2a^2}{24T^2} \left(
    3 (1-8 \xi ) \frac{ K_3\left(x_n\right)}{x_n}
    + \left(\frac{24 \xi }{x_n^2}+1\right) K_2\left(x_n\right)\right) \nn\\
    &\hspace{20em}
    +\frac{n^2\omega^2 (1-4 \xi )}{2T^2}
    \frac{K_2\left(x_n\right)}{x_n^2}
    \Bigg)
    + \cO(\mu_{\mu\nu}^4), \nn\\
    \cP 
    &= \frac{m^4}{\pi ^2}\sum_{n=1}^\infty
    \cosh (\nu  n) \Bigg(
    \frac{K_2\left(x_n\right)}{x_n^2}
    + \frac{n^2a^2}{24T^2} \left(
    12(1-4\xi ) \frac{K_2(x_n)}{x_n^2}
    + (24 \xi -5)\frac{K_3\left(x_n\right)}{x_n}\right)
    + \frac{\xi  n^2\omega^2}{T^2} \frac{K_2\left(x_n\right)}{x_n^2}
    \Bigg)
    + \cO(\mu_{\mu\nu}^4), \nn\\
    \cX_{aa}
    &= \frac{m^4}{\pi ^2}\sum_{n=1}^\infty
    \cosh (\nu  n) 
    \frac{n^2}{4T^2} \left(
    2(2\xi -1) \frac{ K_2\left(x_n\right)}{x_n^2}
    + (1-4 \xi )\frac{ K_3\left(x_n\right)}{x_n}\right)
    + \cO(\mu_{\mu\nu}^2)
    , \nn\\
    \cX_{\omega\omega}
    &= \frac{m^4}{\pi ^2}\sum_{n=1}^\infty
    \cosh (\nu  n) 
    \frac{n^2}{2T^2}
    (2 \xi -1) 
    \frac{K_2\left(x_n\right)}{x_n^2}
    + \cO(\mu_{\mu\nu}^2), \qquad
    \cX_{a\omega} 
    = \cX_{a\ell}
    = \cX_{\omega\ell} = \cO(\mu_{\mu\nu}^2), \nn\\
    \cQ_\ell 
    &= \frac{m^4}{\pi ^2}\sum_{n=1}^\infty
    \cosh (\nu  n) 
    \frac{-n^2}{6T^2} \left(3(2\xi -1) \frac{K_2\left(x_n\right)}{x_n^2}+\frac{K_3\left(x_n\right)}{x_n}\right)
    + \cO(\mu_{\mu\nu}^2), \qquad 
    \cQ_{a}
    = \cQ_{\omega} = \cO(\dow^2), \nn\\
    \cN 
    &= \frac{m^3}{\pi ^2}\sum_{n=1}^\infty
    \sinh (\nu  n) 
    \left(
    \frac{K_2\left(x_n\right)}{x_n}
    + \frac{n^2a^2}{24T^2} 
    \left(K_3\left(x_n\right)-\frac{4 K_2\left(x_n\right)}{x_n}\right) \right)
    + \cO(\mu_{\mu\nu}^4), \nn\\
    \cN_\ell 
    &= \frac{m^3}{\pi ^2}\sum_{n=1}^\infty
    \sinh (\nu  n) \frac{-n^2}{6T^2}\frac{K_2\left(x_n\right)}{x_n}
    + \cO(\mu_{\mu\nu}^2), \qquad
    \cN_a = \cN_\omega = \cO(\mu_{\mu\nu}^2).
\end{align}
The improvement terms to cast these in thermodynamic form are given as
\begin{align}
    \gamma_1 
    &= \Lambda_1
    + \frac{m^2}{12\pi^2}\sum_{n=1}^\infty
    \cosh (\nu  n) 
    \left(4 (1-3 \xi ) \frac{K_1\left(x_n\right)}{x_n}
    + K_0\left(x_n\right)\right)
    + \cO(\mu_{\mu\nu}^2), \nn\\
    \gamma_2
    &= \frac{(1-6\xi)m^2}{6\pi^2}\sum_{n=1}^\infty
    \cosh (\nu  n) 
    \frac{K_1\left(x_n\right)}{x_n}
    + \cO(\mu_{\mu\nu}^2), \nn\\
    \lambda_1 
    &= \frac{\dow}{\dow\mu}\Lambda_1
    + \frac{m}{6\pi^2}\sum_{n=1}^\infty
    \sinh (\nu  n) 
    K_1\left(x_n\right)
    + \cO(\mu_{\mu\nu}^2), \nn\\
    \lambda_2 
    &= \frac{m}{6\pi^2}\sum_{n=1}^\infty
    \sinh (\nu  n) 
    K_1\left(x_n\right)
    + \cO(\mu_{\mu\nu}^2),
\end{align}
for an arbitrary function $\Lambda_1(T,\mu)$.
The resultant EoS is given as
\begin{align}
    \boxed{p = \frac{m^4}{\pi^2}\sum_{n=1}^\infty
    \cosh (\nu  n) 
    \Bigg(
    \frac{K_2\left(x_n\right)}{x_n^2} 
    + \frac{n^2a^2}{24T^2}\frac{K_3\left(x_n\right)}{x_n}\Bigg)
    + a^2 T\df_T\Lambda_1
    + 2 \omega^2 \Lambda_1.}
\end{align}
The quadratic invariant in this case can be computed to be
\begin{align}
    \cI_2 = \frac{m^4}{12\pi^2T^2}\sum_{n=1}^\infty
    \cosh (\nu  n) n^2\frac{K_3\left(x_n\right)}{x_n}.
\end{align}




\end{widetext}

\end{document}